\documentclass{llncs}
\usepackage{frameit,amsmath,color,latexsym,graphics,wrapfig,amssymb}
\usepackage{times,epsfig}
\usepackage{graphicx}

\usepackage{ifthen}
\usepackage{times,frameit,epsfig,amsmath,color,latexsym,graphics,wrapfig,clrscode3e}
\usepackage{frameit,amsmath,color,latexsym,graphics,wrapfig,textcomp,dsfont}
\usepackage{hyperref}
\usepackage{url}
\usepackage{multirow}
\usepackage{bussproofs}

\usepackage{listings}
\definecolor{codegreen}{rgb}{0,0.6,0}
\definecolor{codegray}{rgb}{0.5,0.5,0.5}
\definecolor{codepurple}{rgb}{0.58,0,0.82}
\definecolor{backcolour}{rgb}{1,1,1}

\lstdefinestyle{mystyle}{
	backgroundcolor=\color{backcolour},
	commentstyle=\color{codegreen},
	keywordstyle=\color{magenta},
	numberstyle=\tiny\color{codegray},
	stringstyle=\color{codepurple},
	basicstyle=\ttfamily\scriptsize,
	breakatwhitespace=false,
	breaklines=true,
	captionpos=b,
	keepspaces=true,
	numbers=left,
	numbersep=2pt,
	showspaces=false,
	showstringspaces=false,
	showtabs=false,
	tabsize=2,
	basewidth  = {.5em,0.4em}
}

\newcommand{\setvars}[1]{\ensuremath{\bar{#1}}}

\newcommand{\savespace}{\vspace{-2mm}}
\newcommand{\saveone}{\vspace{-1mm}}

\newcommand{\RA}{\ensuremath{\mathcal{R}}}

\newcommand{\UNK}{\code{U}}
\newcommand{\UNKP}{\code{P}}

\newcommand{\inv}{\myit{invert}}

\newcommand{\flow}{\ensuremath{\tau}}
\newcommand{\okF}{\ensuremath{\surd}}

\newcommand{\lemerr}{\code{err}}

\newcommand{\sepnodeF}[3]{\ensuremath{{#1}{\pto}#2(#3)}}
\newcommand{\sepnode}[3]{\ensuremath{#1{\mapsto}#2(#3)}}
\newcommand{\seppredF}[2]{\ensuremath{#1(#2)}}
\newcommand{\seppred}[2]{\ensuremath{#1(#2)}}

\newcommand{\self}{\btt{root}}
\newcommand{\report}[1]{ }

\newcommand{\acm}[1]{ }

\newcommand{\hide}[1]{}
\newcommand{\hideie}[1]{}
\newcommand{\der}{\ensuremath{\texttt{-\!>}}}
\newcommand{\nil}{\btt{null}}
\newcommand{\res}{\btt{res}}
\newcommand{\emp}{\btt{emp}}

\def\emptyheap{\sc{e}}

\newcommand{\pure}{\ensuremath{\pi}}

\newcommand{\heap}{\ensuremath{\kappa}}

\newcommand{\constr}{\ensuremath{\Phi}}

\newcommand{\sobd}{{\code{SOBD}}}

\newcommand{\entailSA}[5]{\ensuremath{#2~{\vdash}~#3\,{\yields}\,(#4,#5)}}
\newcommand{\entailLEM}[6]{\ensuremath{#1{:}~#2~{\vdash}^{#3}_{#4}~#5\,{\yields}\,#6}}
\newcommand{\entailL}[5]{\ensuremath{#1~{\vdash}_{#3}~#2\,{\yields}\,(#4,#5)}}

\newcommand{\entailKE}[5]{\ensuremath{#3{\vdash}^{#1}_{#2}#4\,{\yields}\,#5}}

\newcommand{\entailVVE}[3]{\entailKE{\heap}{\code{V}}{#1}{#2}{#3}}

\newcommand{\entailSYN}[5]{{#1}\vdash_{#2}{#3} ~{\yields}~({#4},{#5})}
\newcommand{\myit}[1]{\textit{#1}}

\newcommand{\FIL}[2]{\ensuremath{{\bigtriangledown}(#1,#2)}}
\newcommand{\project}[2]{\ensuremath{{\Pi}(#1,#2)}}

\newcommand{\chain}[3]{\ensuremath{{\code{h{chain}}}(#1,#2,#3)}}

\def\sep{\code{*}}

\newcommand{\entrulen}[1]{[\underline{{\bf \scriptstyle #1}}]}

\def\FV{\myit{FV}}

\def\unfold{\btt{unfold}}

\def\lemapp{\btt{lem\_app}}

\def\D{\Delta}

\def\fresh{\myit{fresh}}

\def\int{\code{int}}

\def\true{\code{true}\,}
\def\false{\code{false}\,}

\newcommand{\hash}{\ensuremath{{\scriptstyle \#}}}
\newcommand{\num}[1]{|#1|}

\newcommand{\seph}[2]{\ensuremath{#1{\mapsto}#2}}
\newcommand{\nseph}[1]{\ensuremath{#1{\nptto}{\_}}}
\newcommand{\nptto}{\ensuremath{\not\mapsto}}
\newcommand{\consh}[1]{\ensuremath{\Gamma^h(#1)}}
\newcommand{\consd}[1]{\ensuremath{\Gamma^d(#1)}}
\newcommand{\cons}[1]{\ensuremath{\Gamma(#1)}}
\newcommand{\rcons}[1]{\ensuremath{\overline{\Gamma}(#1)}}
\newcommand{\rpto}{\ensuremath{\overline{\eta}}}

\newcommand{\rconsh}[1]{\ensuremath{\overline{\Gamma}^h(#1)}}
\newcommand{\rconsd}[1]{\ensuremath{\overline{\Gamma}^d(#1)}}
\newcommand{\Varh}[1]{\ensuremath{\Vs^h(#1)}}
\newcommand{\Vs}{\ensuremath{\mathcal{V}}}
\newcommand{\Vard}[1]{\ensuremath{\Vs^d(#1)}}
\newcommand{\lem}{\ensuremath{\iota}}
\newcommand{\lemstore}{\ensuremath{{\sc L}}}

\newcommand{\code}[1]{{\small {\ensuremath{\tt #1}}}}
\newcommand{\sm}[1]{{\small \mbox{$#1$}}}
\newcommand{\btt}[1]{{\ensuremath{\tt #1}}}

\newcommand{\nodo}[1]{}

\newcommand{\defpred}{\btt{pred}}

\newcommand{\head}{\btt{R}}

\newcommand{\ensures}{\btt{ensures}}
\newcommand{\requires}{\btt{requires}}

\newcommand{\synlemma}[5]{\btt{syn}({#1},{#2},{#3},{#4},{#5})}

\newtheorem{defn}{Definition}

\def\G{\Gamma}
\def\S{\Sigma}

\def\G{\Gamma}

\def\sep{\ensuremath{*}}
\newcommand{\pto}{{\scriptsize\ensuremath{\mapsto}}}
\newcommand\eq[3]{\ensuremath{\myit{eq}_{#1}(#2,#3)}}

\newcommand{\Deq}[2]{\ensuremath{\frac{\begin{array}{c}#1\end{array}}{\begin{array}{c}#2\end{array}}}}

\def\Var{\myit{Var}}

\def\inv{\bigtriangledown}

\def\invo{\overline{\bigtriangledown}}

\def\fresh{\myit{fresh}}

\newcommand{\atom}{\alpha}
\newcommand{\reduce}{\sigma}

\newcommand{\subst}[2]{\ensuremath{[#1 {/} #2]}}
\newcommand{\yields}{\leadsto}

\newcommand{\anon}{\ensuremath{\_\,}}

\newcommand{\satp}[1]{\bf \footnotesize{\scriptstyle{sat}}(\ensuremath{#1})}
\newcommand{\pureentail}[3]{\ensuremath{#1 ~{\implies}~ #2 ~{\yields}~ #3}}

\newcommand{\ent}{\ensuremath{{\vdash}}}
\newcommand{\imply}{\ensuremath{\Rightarrow}}
\newcommand{\implyeq}{\ensuremath{\equiv}}

\newcommand{\form}[1]{\ensuremath{#1}}

\newcommand{\horn}{\ensuremath{\sigma}}
\newcommand{\Horn}{\RA\hide{\mathcal{C}}}

\newcommand{\segpred}{{\bf SPred}}

\newcommand{\ass}{\ensuremath{\longrightarrow}}

\newcommand{\toolname}{\code{S}}

\newcommand{\cyclic}{\code{Cyclic}$_{SL}$}
\newcommand{\ud}{{\code{UD}}}
\newcommand{\baga}{{\sc{\scriptstyle BagA}}}

\newcommand{\xpure}{{\bf {\scriptstyle XPure}}}

\newcommand{\NI}{{\sm\hash}}

\newcommand{\llem}{\ensuremath{\rightarrow}}
\newcommand{\rlem}{\ensuremath{\leftarrow}}

\def\qed{\hfill\ensuremath{\square}}

\def\assumpset{\ensuremath{{\cal R}}}

\newcommand{\conferencepaper}{1} 
\newcommand{\rep}[1]{\ifthenelse{\conferencepaper = 0}{#1}{}}
\newcommand{\conf}[1]{\ifthenelse{\conferencepaper = 0}{}{#1}}
\newcommand{\repconf}[2]{\ifthenelse{\conferencepaper = 0}{#1}{#2}}
\begin{document}
 \title{Enhancing Inductive Entailment Proofs in Separation Logic with Lemma Synthesis}



\author{Quang Loc Le \thanks{This work was done in part when the author was a PhD student in National University of Singapore.}
}

\institute{
 Teesside University, UK
}

\maketitle

\begin{abstract}
This paper presents an approach to lemma synthesis to support
advanced inductive entailment procedures based on separation logic.
We first propose a
mechanism where lemmas are automatically proven and  systematically applied.
Our lemmas may include  {\em universal} guard and/or {\em unknown} 
predicate.
While the former is critical for expressivity,
the latter is essential for
supporting relationships between multiple predicates.
We further introduce lemma synthesis to support
(i) automated inductive reasoning together with
frame inference
 and (ii) theorem exploration.
For (i) we automatically discover and prove
auxiliary lemmas during an inductive proof; and for (ii)  we automatically
 generate a useful set of lemmas to relate user-defined 
or system-generated predicates.
We have implemented our proposed approach into an existing  verification system
and tested its capability in inductive reasoning and theorem exploration.
The experimental results show that the enhanced 
system can automatically synthesize useful lemmas to facilitate reasoning on
 a broad range of non-trivial inductive proofs.

\keywords Lemma Synthesis, Induction Proving, Theorem Exploration,
 Separation Logic, User-Defined Predicate.

\end{abstract}

\section{Introduction}
\label{sec:intro}
 Separation logic (SL) \cite{Ishtiaq:POPL01,Reynolds:LICS02}
 has been well established for reasoning
 about heap-based
 programs.
Frame rule in SL enables
  modular (compositional) reasoning
in the presence of the heap
and is essential for scalability \cite{Jacobs:NFM:2011,CAV08:Yang,Cristiano:POPL:09}.
In the last decade,
 a large number of proof systems for SL
  have been studied
 \cite{Berdine:APLAS05,Brotherston:CADE:11,Cristiano:POPL:09,DBLP:conf/cade/IosifRS13,Chin:SCP:12,Madhusudan:POPL:2012}. 
Generally speaking, the key challenges of these systems are
  to support 
bi-abduction (automated frame inference \cite{Berdine:APLAS05,Chin:SCP:12} and
 logical abduction \cite{Cristiano:POPL:09,Loc:CAV:2014,Trinh:APLAS:2013}),
and automated induction proving \cite{Brotherston:CADE:11}
in SL fragments with inductive predicates.
While the use of general inductive predicates attains expressive power,
a powerful entailment procedure supporting the inductive predicates needs to  meet   
the following two main challenges.

{\bf Induction Reasoning.}
Entailment checks involving inductive predicates normally require
induction.
For an indirect solution,
existing works employed
lemmas that are consequences of induction.
These lemmas were either hardwired for a set of predefined predicates
(i.e. lists \cite{Berdine:APLAS05}),
or automatically generated (for normalization of shape analysis \cite{Loc:CAV:2014},
or for some predefined classes of predicates \cite{EneaSW:ATVA:15}).
 While these lemma approaches can handle
 induction reasoning for some specific predicates,
they could not support induction in general proofs.
%
 Brotherston et. al. make an important step
towards automating general induction reasoning in SL with cyclic entailment proofs \cite{Brotherston:CADE:11,Brotherston:APLAS:12}.
Recently, authors in \cite{Chu:PLDI:2015} managed induction by a 
framework with historical proofs.
However, these proof systems \cite{Brotherston:CADE:11,Chu:PLDI:2015}
 did not consider
frame inference.
Thus, they
 provide limited support
for the frame rule 
as well as {\em modular} verification.


{\bf Completeness.} Past works 
introduce decision procedures for SL decidable fragments
including
 hardwired lists \cite{Berdine:APLAS05,Bouajjani:ATVA:2012,Cook:CONCUR:2011,pldi:PerezR11}, or even
user-defined predicates with some syntactic restrictions \cite{DBLP:conf/cade/IosifRS13}.
However, the SL fragment with (arbitrary) user-defined predicates
(and arithmetic constraints) is, in general, undecidable.
Entailment procedures for this general fragment
typically trade off completeness for expressiveness \cite{Brotherston:CADE:11,Chin:SCP:12,Distefano:2008:OOPSLA,Madhusudan:POPL:2012}.
To enhance the completeness of program verification,
there have been efforts of exploring relations between predicates via {\em user-supplied}
 lemmas 
 \cite{Berdine:APLAS05,Brotherston:CADE:11,Brotherston:APLAS:12,Distefano:2008:OOPSLA,Jacobs:NFM:2011,Nguyen:CAV08}.
While such a static approach puts creative control back into the users' hands, it is not fully automatic
and is infeasible to support inductive proofs relying on
auxiliary
lemmas of 
dynamically synthesized predicates (like those in \cite{Brotherston-Gorogiannis:SAS:14,Loc:CAV:2014}).

In this work, we propose an approach to lemma synthesis
for 
advanced inductive proofs in a SL fragment with user-defined predicates
and Presburger arithmetic.
Our technical starting point is an entailment procedure
 for user-defined predicates (i.e. those procedures
in the spirit of \cite{Chin:SCP:12})
combined with second-order bi-abduction \cite{Loc:CAV:2014}.
We extend 
 this proof system with a new mechanism
where lemmas are automatically generated, proven
 and systematically applied.
Finally, we apply lemma synthesis
 into
 theorem exploration.

 Frame inference has been studied in SL entailment procedures like \cite{Berdine:APLAS05,Chin:SCP:12,Piskac:CAV:2013}. Intuitively, these systems prove
 validity of an entailment \form{\D_a \ent \D_c} by unfolding user-defined predicates, subtracting heap predicates
until
 halting at sequents
 with {\em empty} heap (i.e. the \form{\emp} predicate in SL) in the consequent,
 such that \form{\D_{\code{f}} {\ent} \textcolor{blue}{{\emp}}{\wedge}\pure_r} and then,
 conclude the entailment is valid with the frame \form{\D_{\code{f}}}, denoted by \form{\D_{a} {\ent}\D_c \yields \D_{f}}.
 However, these systems did not provide a direct solution for general induction proofs. 
We tackle this challenging of
frame inference for inductive entailment proofs via the new lemma systhesis.
Our key insight is that the entailment check \form{\D_a {\ent} \D_c {\yields}\D_f}
is semantically
equivalent to the check \form{\D_a {\ent} \D_c {\sep}\D_f {\yields}\emp}.
To infer frame for inductive entailment checks like the former, we will
prove the latter check inductively while inferring frame \form{\D_f} abductively.
Concretely, we assume frame as an unknown predicate \code{U}, construct
the conjecture \form{\code{lemma~l_1{:}}\D_a {\rightarrow} \D_c {\sep} \code{U}}, and
finally inductively prove this lemma and abductively 
infer a definition of the predicate \code{U}.
The benefit of the use of lemma synthesis in our approach is twofold.
First, our proposed approach is easily integrated into existing proof systems
with lemma mechanism i.e. \cite{Nguyen:CAV08,Distefano:2008:OOPSLA}. 
Second, the synthesized lemmas (i.e. \code{l_1}) are accumulated for reuse in future.

 Lemmas in our system may include
universally quantified guards  
 and unknown predicates.
We will use the notation \form{\D[\setvars{v}]} to stand for a formula
with free variables \form{\setvars{v}}.
Our lemmas with universal guards
\footnote{
We will refer lemma with universal guard as universal lemma.}
 have the form \form{\forall \setvars{v} \cdot H(\setvars{v}) {\wedge} G(\setvars{v}) \rightarrow B(\setvars{v})}
whereas universal guard \form{G} and body \form{B} may include unknown predicates whose definitions
 need to be inferred.
While guards over universal variables make our mechanism very expressive,
 unknown predicates enable us to synthesize generic lemmas i.e. those with weak(est) guards
and strong(est) bodies.
The meaning of lemmas is interpreted in classic semantics \cite{Ishtiaq:POPL01} i.e. for the lemma above \form{\forall \setvars{v} \cdot H(\setvars{v}) {\wedge} G(\setvars{v}) \models B(\setvars{v})}; the LHS exactly entails the RHS (with empty heap in residue).
Our lemma proving is based on the principle of cyclic proof \cite{Brotherston:CADE:11,Brotherston:APLAS:12},
and can support induction proving.
For universal lemma synthesis, given the entailment check
 \form{\exists \setvars{e} {\cdot} \D_a{\wedge}G(\setvars{e})  ~{\ent}~ \D_c}, our system would generate the lemma
\code{lemma ~ l_2{:}} \form{\forall \setvars{e} {\cdot} \D_a {\wedge} \code{P}(\setvars{e})  ~ {\rightarrow}~ \D_c}, whereas \code{P}
is a newly-inferred predicate. 



Our predicate inference mechanism is based on
 the principle of second-order bi-abduction ({\sobd}), for shape domain \cite{Loc:CAV:2014} and for pure (non-shape)
domains \cite{Trinh:APLAS:2013},
which is an extension of bi-abduction \cite{Cristiano:POPL:09} to user-defined predicates.
 A {\sobd} 
entailment procedure takes two SL formulas \form{\D_{ante}} and \form{\D_{conseq}}
as inputs. It infers missing hypotheses \form{\RA?}
and residual frame \form{\D_{frame}?}. 
\form{\RA?} is a set of Horn clause-based constraints 
 over unknown
variables of \form{\D_{ante}} and \form{\D_{conseq}}.
These constraints
 have the form of logical implication
 i.e. \form{\D_L {\imply} \D_R}.
 The set \form{\RA?} can be solved
to obtain definitions of the unknown predicates by the 
algorithms in \cite{Loc:CAV:2014,Trinh:APLAS:2013}.

In inductive reasoning, theory exploration is a
 technique for automatically generating
 and proving useful lemmas for sets of given functions, constants and datatypes  \cite{Claessen:2013:CADE,Johansson:2011:AR,Roy:2007}.
In our context, theory exploration is meant for discovering
useful lemmas for a set of given user-defined predicates.
In SL, this technique was indeed presented 
\cite{Berdine:APLAS05,EneaSW:ATVA:15}.
Like \cite{EneaSW:ATVA:15} and unlike
\cite{Berdine:APLAS05}, our approach is automation-based. 
Different from these, 
our technique is capable of generating lemmas with newly-inferred user-defined predicates
 i.e. those that are synthesized while proving the lemmas.

The novelty of our proof system is the lemma synthesis
with predicate inference. 
Our primary contributions are summarized as follows:
\begin{itemize}\savespace
\item We propose a new mechanism 
where 
 universal lemmas 
are soundly synthesized and systematically applied.
\item We synthesize lemmas to support  inductive proofs together with frame inference and theorem exploration.
\item We have implemented the proposal in an entailment procedure, called {\toolname},
and integrated {\toolname} into a modular verification.
Our experiments on
sophisticated inductive proofs show that our approach is promising for advancing
state-of-the-art in automated verification of heap-manipulating programs.
\end{itemize}


\section{Motivation and Overview}
\label{sec:motivate}

\subsection{Entailment Procedure using Universal Lemma Synthesis}
We extend an entailment procedure
with basic inference rules
(i.e. \cite{Brotherston:CADE:11,Chin:SCP:12})
to inference capability
with the second-order bi-abduction ({\sobd}) mechanism \cite{Loc:CAV:2014} 
and a new lemma mechanism.
This lemma mechanism enhances the inference rules
with a set of external and proven lemmas \form{\lemstore}.
These lemmas can be initially supplied by users
as well as additionally and dynamically synthesized while the entailment is proven.
The enhanced entailment procedure is formalized as follows:
\form{
\entailL{\D_{\code{ante}}}
{\D_{\code{conseq}}}{\lemstore} {\RA}{\D_{\code{frame}}}}
 such that:
\form{\lemstore{\wedge}\RA{\wedge} {\D_{ante}} ~{\models}~ {\D_{conseq}} {\sep} \D_{frame}}.
Consequently, inference rules
of the starting system are using the augmented lemmas  \code{\lemstore} in
the new system as the following:\savespace
\[
\begin{array}{l}
\quad \quad \frac{
\entailL{\D_{l_1}}
{\D_{r_1}}{} {\RA_1}{\D_{\code{R_1}}}
 \quad ...\quad
\entailL{\D_{l_i}}
{\D_{r_i}}{} {\RA_i}{\D_{\code{R_i}}}
}
{
\entailL{\D_{1}}
{\D_{2}}{} {\RA}{\D_{\code{R}}}
}
\\
\implies ~
\frac{
\entailL{\D_{l_1}}
{\D_{r_1}}{\lemstore} {\RA_1}{\D_{\code{R_1}}}
 \quad ...\quad
\entailL{\D_{l_i}}
{\D_{r_i}}{\lemstore} {\RA_i}{\D_{\code{R_i}}}
}
{
\entailL{\D_{1}}
{\D_{2}}{\lemstore} {\RA}{\D_{\code{R}}}
}
\end{array}
\]
For a standard proof system for SL with inductive predicates,
please refer to \cite{Chin:SCP:12}. (A summary is presented
in App. \ref{entail.rules}.)
 We shall propose
 \form{\entrulen{LAPP}} and \form{\entrulen{LAPP-\forall}} rules
 for lemma application (Sec. \ref{lem.app}), \form{\entrulen{LSYN}}
rule for lemma synthesis,
\form{\entrulen{R{\sep}}} rule for heap split,
and \form{\entrulen{AU}}, \form{\entrulen{AF}},
 \form{\entrulen{AU-P}}, \form{\entrulen{AF-P}}  rules
 for predicate inference (Sec. \ref{ent.lemsyn}).


To illustrate how our proof system can support
induction reasoning together with
complex frame inference,
consider the following entailment check \code{VC_0}
\savespace\[\savespace
\form{\form{\seppred{\code{ll\_last}}{t{,}t'}{\sep} \seppred{\code{ll}}{y}}}
~ {\ent_{\emptyset}}~
 \sepnode{\code{t'}}{c_1}{q}
\] \savespace
\savespace
\[\savespace
\begin{array}{l}
\form{\defpred~ \seppred{\code{ll}}{\self} {\equiv} \emp {\wedge} \code{\self}{=}\nil }
\form{~ {\vee} ~\exists q {\cdot} \sepnode{\code{\self}}{c_1}{q} {\sep}
\seppred{\code{ll}}{q} } \\
\form{\defpred~\seppred{\code{ll\_last}}{\self{,}s}} {\equiv}
 \form{\sepnode{\code{\self}}{c_1}{\nil}{\wedge}\code{\self}{=}s} 
~ {\vee} \form{~\exists q {\cdot} \sepnode{\code{\self}}{c_1}{q} {\sep}
\seppred{\code{ll\_last}}{q{,}s}}; \\
 \end{array}
\]
whereas  \form{\code{struct~ c_1\{{c_1{\sep} ~ next};\}}}.
\code{VC_0} is a verification condition generated to
verify memory safety of programs
that access the last element
(i.e. \code{g\_list\_append} in \code{glist.c}
 of GLIB library \cite{glib:13}
 - see App. \ref{mov.veri.simp}).
In the predicate definitions above, we use basic SL notations
to express heaps, e.g. 
 empty predicate (i.e. \form{\emp}),
points-to predicate (i.e. $\sepnodeF{\self}{c_1}{p}$
  asserts a concrete heap cell bound with
 an allocated data type \code{c_1}, pointed-to by
 the variable \form{\self} and linked with
 downstream pointer \form{p} via the field \code{next}).

To infer frame for this inductive entailment check,
we assume the frame be a unknown predicate, i.e. \form{ \code{U_2}(t{,}t'\NI{,}q{,}y)}, and form the following conjecture:
\savespace\[\savespace
\code{lemma ~c{:}~}
\form{\seppred{\code{ll\_last}}{t{,}t'}{\sep} \seppred{\code{ll}}{y}} \rightarrow
\sepnode{\code{t'}}{c_1}{q} {\sep} \code{U_2}(t{,}t'\NI{,}q{,}y)
\]
Then, we prove its validity and 
infer a set of relational assumptions as:
\savespace\[\savespace
\begin{array}{l}
\form{\horn_1{:}}~ \form{{\sep} \seppred{\code{ll}}{y}{\wedge}t'{=}t {\wedge}q{=}\nil}~
 \form{~{\imply}~{\code{U_2}(t{,}t'{,}q{,}y)}}\\
\form{\horn_2{:}}~ \form{\sepnode{t}{\code{c_1}}{q_1}
{\sep}\code{U_2}(q_1{,}t'{,}q_2{,}y)}
 \form{~{\imply}~{\code{U_2}(t{,}t'{,}q_2{,}y)}}\\
\end{array}
\]
From \form{\horn_1} and \form{\horn_2}, we
 synthesize
the following definition for \code{U_2}
\savespace\[\savespace
\begin{array}{l}
\form{\seppred{\code{U_2}}{\self{,}t'{,}q{,}y}} {\equiv}
 \form{\seppred{\code{ll}}{y}{\wedge}\code{\self}{=}t' {\wedge}q{=}\nil} 
~ {\vee}~ \form{~\exists ~q_1 {\cdot}~ \sepnode{\code{\self}}{c_1}{q_1} {\sep}
\seppred{\code{U_2}}{q_1{,}t'{,}q{,}y}}; \\
\end{array}
\]
Finally, using theorem exploration presented in Sec \ref{sec:explore}
( and App. \ref{app.pred.trans}),
we generate the following two-way {\em separating} lemmas
 \footnote{We refer \form{A{\leftrightarrow}B} as two-way lemma,
 a short form
of two reverse lemmas: \form{A{\rightarrow}B} and \form{B{\rightarrow}A}.}
 to normalize the predicate \code{U_2}:
\savespace\[\savespace
\begin{array}{l}
\code{lemma ~conseq_0{:}}
\form{\seppred{\code{U_2}}{\self{,}t'{,}q{,}y} \leftrightarrow
\code{U_3}(\self{,}t'){\sep}\seppred{\code{ll}}{y}{\wedge}q{=}\nil} \\
\form{\seppred{\code{U_3}}{\self{,}t'}} {\equiv}
 \form{ \emp{\wedge}\code{\self}{=}t'} 
~ {\vee}~ \form{~\exists ~q_1 {\cdot}~ \sepnode{\code{\self}}{c_1}{q_1} {\sep}
\seppred{\code{U_3}}{q_1{,}t'}}; \\
\end{array}
\]
To sum up, our system successfully proves and derives
  \form{\code{U_3}(t{,}t'){\sep}\seppred{\code{ll}}{y}{\wedge}q{=}\nil}
as frame of \code{VC_0}.
We present an example for universal lemma synthesis in App. \ref{mov-pure}.

\subsection{Modular Verification with Lemma Synthesis}\label{mov-frame}
 \begin{wrapfigure}{l}{0.42\textwidth}\savespace
\begin{center} 
\savespace
  \savespace
 \[
\begin{array}{ll}
1 & \code{void ~check(struc~c_2{\sep}~a)} \\
2 & \code{\{~ while(a{\der}val{==}1)a{=}a{\der}next;} \\
3 & \code{~~~ while(a{\der}val{==}2)a{=}a{\der}next;} \\
4 & \code{~~~ assert~a{\der}val{==}3;} \\
5 & \code{~~~ return};\}
\end{array}
\]  
\savespace \savespace
 \savespace \savespace
\caption{Code of method \code{check}.}
\label{fig.mov.check}
\end{center}\savespace \saveone
\end{wrapfigure}

 In modular verification of heap-based programs,
specifications of recursive methods and loop invariants normally relate
to recursive predicates; consequently, compositionally verifying these specifications
  requires both
 induction reasoning and frame inference. Our proposed approach 
 brings the best support for
such verification.

Consider the code fragment
in Fig. \ref{fig.mov.check}. This code fragment 
 checks whether list segment
pointed by \code{a} is decomposed into three regions:
 a list segment of 1 values (the \code{while} loop at line 2),
 a list segment of 2 values (the \code{while} loop at line 3),
and a 3-value node (the assertion at line 4).
We assume that
the method \code{check} (\code{while} loop at line 2, line 3)
 has been supplied with
the \form{s_1} (\form{s_2},  and \form{s_3}
resp.) specification as:
\begin{small}
\[\begin{array}{l}
\form{(s_1){~} \requires ~\seppred{\code{ls1}}{a{,}p_1} {\sep}\seppred{\code{ls2}}{p_1{,}p_2}{\sep} \sepnode{p_2}{\code{c_2}}{3,\nil}~ \ensures~ \true;} \\
\form{(s_2){~}\requires ~\seppred{\code{ls1}}{a{,}p_3}{\sep}\sepnode{p_3}{\code{c_2}}{v_1{,}p_4}{\wedge}v_1{\neq}1} 
\form{~~ \ensures~ \seppred{\code{ls1}}{a{,}p_3}{\sep}\sepnode{p_3}{\code{c_2}}{v_1{,}p_4}{\wedge}a'{=}p_3;}\\
\form{(s_3){~}\requires ~\seppred{\code{ls2}}{a{,}p_5}{\sep}\sepnode{p_5}{\code{c_2}}{v_2{,}p_6}{\wedge}v_2{\neq}2} 
\form{~~ \ensures~ \seppred{\code{ls2}}{a{,}p_5}{\sep}\sepnode{p_5}{\code{c_2}}{v_2{,}p_6}{\wedge}a'{=}p_5;}\\
\end{array}
\]
\end{small}
 whereas \form{a'} is value of a after the loop;
 the data structure and the predicates \code{ls1}, \code{ls2} are defined as:
\code{struct ~c_2\{int~val;struct~c_2{\sep}~next;\}}
\begin{small}
 \[
\begin{array}{lr}
\begin{array}{l}
\form{\defpred~ \seppred{\code{ls1}}{\self{,}s}} {\equiv}
\form{\emp} {\wedge} \form{\code{\self}{=}s } \\
~ {\vee} \form{~\exists ~q {\cdot}~ \sepnode{\code{\self}}{\code{c_2}}{1,q} {\sep}
\seppred{\code{ls1}}{q{,}s} };  \\
\end{array}
 \quad& \quad
\begin{array}{l}
\form{\defpred~\seppred{\code{ls2}}{\self{,}s}} {\equiv}
\form{\emp} {\wedge} \form{\code{\self}{=}s } \\
~ {\vee} \form{~\exists q {\cdot}~ \sepnode{\code{\self}}{\code{c_2}}{2{,}q} {\sep}
\seppred{\code{ls2}}{q{,}s}}; \\
\end{array}
\end{array}
\]
\end{small}
As a (bottom-up) modular verification,
the loops are verified prior to
the verification of the method \code{check}; and the correctness of a method
is reduced to the validity of appropriate verification conditions generated.
Our system generates verification conditions to ensure absence of memory
errors (no null dereference, no double free and no memory leak),
 validity of  functional calls/loops via  compositional pre-/post- conditions
 and post-conditions holding.
 For illustrating the proposed approach,
we briefly discuss the reasoning on the verification condition (\code{VC_1})
 that was generated before 
the loop at line 2 (to prove that the current context can imply the loop invariant):

\[
\form{\seppred{\code{ls1}}{a{,}p_1} {\sep}\seppred{\code{ls2}}{p_1{,}p_2}{\sep} \sepnode{p_2}{\code{c_2}}{3,\nil}
~{\ent}_{\emptyset}~\seppred{\code{ls1}}{a{,}p_3}{\sep}\sepnode{p_3}{\code{c_2}}{v_1{,}p_4}{\wedge}v_1{\neq}1}
\]

(The lemma store \form{\lemstore{=}\emptyset} means there are no user-supplied lemmas.)
\code{VC_1} requires both induction reasoning (for the \form{\code{ls1}{a{,}p_1}} predicate)
and frame inference to prove safety of the rest of the program.
Hence, this entailment is beyond
the capability of most existing SL verification systems (like \cite{Nguyen:CAV08,Brotherston:CADE:11,Chin:SCP:12,Madhusudan:POPL:2012,Piskac:CAV:2013}).

 Instead of instantiating and 
deducing a frame like \cite{Berdine:APLAS05,Chin:SCP:12}, we
assume frame as an unknown predicate and infer this predicate
via abduction.
This inference was implemented in the proposed inference rule
\form{\entrulen{LSYN}}. 
 Concretely, the proof of \code{VC_1} is as follows:
\[
\begin{array}[t]{c}
\code{lemsyn}(\D_a,\D_c,\emptyset) {\yields} (\code{lemma ~l_3{:}}~\form{\D_l \rightarrow \D_r{\sep} \code{U}(\self{,}p_1{,}p_2{,}p_3{,}p_4{,}v_1)}{,}\true )
\\
\form{\seppred{\code{ls1}}{a{,}p_1} {\sep}\seppred{\code{ls2}}{p_1{,}p_2}{\sep} \sepnode{p_2}{\code{c_2}}{3,\nil}  ~ {\ent}_{\emptyset}~ \D_l[a/\self]
~{\yields}~\emp{\wedge}\true}
\\
\hline
\begin{array}[t]{l}
\form{\seppred{\code{ls1}}{a{,}p_1} {\sep}\seppred{\code{ls2}}{p_1{,}p_2}{\sep} \sepnode{p_2}{\code{c_2}}{3,\nil}}\\
\form{\quad ~{\ent}_{\emptyset}~  \seppred{\code{ls1}}{a{,}p_3}{\sep}\sepnode{p_3}{\code{c_2}}{v_1{,}p_4}{\sep} 
 \textcolor{blue}{\code{U}(\setvars{v})}{\wedge}v_1{\neq}1~ {\yields}~ (\true{,}\code{U}(\setvars{v}))}
\end{array}
\end{array}
\]
whereas
\form{\code{U}(\setvars{w}){\equiv}\code{U}(a\NI{,}p_1{,}p_2{,}p_3\NI{,}p_4{,}v_1)}.
 The lemma \code{l_3} was 
 synthesized as
\[
\begin{array}{l}
\code{lemma ~l_3{:}}~\form{\seppred{\code{ls1}}{\self{,}p_1} {\sep}\seppred{\code{ls2}}{p_1{,}p_2}{\sep}
 \sepnode{p_2}{\code{c_2}}{3,\nil}}\\
\qquad \form{~{\rightarrow}~\seppred{\code{ls1}}{\self{,}p_3}{\sep}\sepnode{p_3}{\code{c_2}}{v_1{,}p_4}
{\sep}\code{U}(\self{,}p_1{,}p_2{,}p_3{,}p_4{,}v_1){\wedge}v_1{\neq}1}
\end{array}
\]
\[
\begin{array}{l}
\form{\defpred ~\seppred{\code{U}}{a{,}p_1{,}p_2{,}p_3{,}\self{,}v_1}} ~{\equiv}~ \form{\self{=}\nil {\wedge} {a{=}p_1} {\wedge}p_1{=}p_2 {\wedge}{a{=}p_3}{\wedge}v_1{=}3}\\
\form{\quad \vee~ \sepnode{\self}{\code{c_2}}{v_1',v_4'}{\sep}\seppred{\code{U}}{a{,}p_1{,}p_2{,}p_3{,}v_4'{,}v_1'} {\wedge}
 {a{=}p_1} {\wedge}{a{=}p_3}{\wedge}v_1{=}2{\wedge}v_1'{\neq}1}
\end{array}
\]
The proof derived for the lemma \code{l_3} will be presented in detail in Sec \ref{sec.lemprove.examp}.


For automation, we integrated the proposed
proof system into S2 \cite{Loc:CAV:2014}, a specification inference system.
Simultaneously, we enhanced S2 beyond the shape domain.
We show how the proposed lemma synthesis
was integrated into
 a modular verification with incremental specification
inference over
shape and size properties in App. \ref{mov.veri}.


\section{Preliminaries}
\label{sec:prm}
\noindent{\bf A Fragment of Separation Logic. }
The syntax of the fragment is 
as follows.
\[
\begin{array}{c}
\begin{array}{l}
\constr ~{::=}~ \D ~|~ \constr_1 ~{\vee}~ \constr_2
\quad
  \D ~{::=}~ \exists \bar{v}{\cdot}~(\heap{\wedge}\pure)
 \quad
 \heap ~{::=}~ \emp ~|~ 
\sepnodeF{r}{c}{\setvars{t}} ~|~
\seppredF{\code{P}}{\setvars{t}}  ~|~
\seppredF{\code{U}}{\setvars{t}}
~|~\heap_1 {\sep}\heap_2 \\
\hide{&&~~~~~|~\heap_1 {\sep}\heap_2
~|~ \heap_1
{\magicW}\heap_2\\}
\hide{&&~~~~~|~ \textcolor{red}{\heap_1
{\wedge}\heap_2}\quad}
 \pure ~{::=}~ 
\pure {\wedge} \phi \mid \phi \mid p(\setvars{t})\qquad
 \phi ~{::=}~  \atom
 \mid  \myit{i} \mid
  \exists v{\cdot}~ \phi ~|~ \forall v{\cdot}~ \phi  ~|~ \neg \phi
 ~|~ \phi_1{\wedge}\phi_2 ~|~ \phi_1{\vee}\phi_2 \\
 \end{array} \\
\begin{array}{l}
\defpred ~\seppred{\code{P}}{\setvars{v}}{\equiv}\constr(\setvars{v}) \qquad
 c \in {\myit{Data Types}} \qquad 
  t_i,v, r\in \text{\Var} \qquad
  \setvars{t}\equiv t_1,{\ldots},t_n 
\end{array}
\end{array}
\]

A formula $\constr$ can be a disjunctive formula (\form{\constr_1 ~{\vee}~ \constr_2}).
A conjunctive formula $\D$ is conjoined by
 spatial formula \form{\heap} and pure formula \form{\pure}.
All free variables 
are implicitly universally quantified
at the outermost level.
\form{\nil} is a special heap location.
In the predicate \form{\sepnodeF{r}{c}{\setvars{t}}}, \form{r} is a root variable.
For abduction reasoning, our fragment also includes
{\em unknown predicates}, spatial (\form{\seppredF{\code{U}}{\setvars{t}}})
and pure (\form{p(\setvars{t})}) second-order variables, whose definitions need to be inferred \cite{Loc:CAV:2014,Trinh:APLAS:2013}.
Pure formulas are 
 constraints over (in)equality \form{\atom} (on pointers), and
Presburger arithmetic \form{i}. 
Note that \form{v_1 {\neq} v_2}
and \form{v {\neq} \nil} are short forms for \form{\neg (v_1{=}v_2)}
 and \form{\neg(v{=}\nil)}, respectively.
 We occasionally use a sequence (i.e. \form{\setvars{t}})
to denote a set when it is not ambiguous. We omit \form{\pure} when it is \form{\true}.
A formula without any {user-defined} predicate instances is referred as a base formula.
\hide{
noindent {\bf Notation: }
we define the following notations to be used in the rest of the paper:
\begin{itemize}
\item \form{\consh{\D}} is a set of constraints on pointer variables of \form{\D}.
\item \form{\consd{\D}} is a set of data constraints of \form{\D}.
\item \form{\pto(\D)} is a set of point-to formulas of \form{\D}.
\end{itemize}

Generally, given a set of constraints \form{\cons{\D}}, we use \form{\rcons{\D}} to denote
a conjunction of the constraints on this set i.e.
\[
\rcons{\D} = \bigwedge \{\ass ~|~ \ass \in \cons{\D}  \}
\]
Similarly, we use \form{\rpto({\D})} to denote
a spatial conjunction of the point-to relations i.e.
\[
\rpto({\D}) = \sep \{\seph{v}{E} ~|~ \seph{v}{E} \in \pto({\D})  \}
\]
Finally, a conjunction formula can be expressed:
\[
 \D = \rpto(\D) \wedge \rconsh{\D} \wedge  \rconsd{\D}
\]

\Varh{\D} is the set of pointer variables in the formula \form{\D}.
\Vard{\D} is the set of non-pointer variables in the formula \form{\D}.
For example, 
\[
\begin{array}{l}
\Varh{\seph{v_h}{v_d}} = \{v_h\} \\
\Vard{\seph{v_h}{v_d}} = \{v_d\} \\
\Varh{\nseph{v_h}} = \{v_h\} \\
\Vard{\nseph{v_h}} = \{~\} \\
\end{array}
\]

We would like to note that if the context about the formula is clear,
we will use the notations without any parameter.
}


\noindent{\bf User-Defined ({\ud}) Predicate. }
A {\ud} predicate \code{P} is defined as
\form{
\defpred~ \seppred{\code{P}}{\setvars{v}}  {\equiv}
   \bigvee^n_{i=1} ({\exists} \setvars{w_i}{\cdot}~ \D_i );
}
whereas \code{P} is predicate name. 
\form{\setvars{v}} is a set of formal parameters. 
 \form{\bigvee^n_{i=1} {\exists} \setvars{w_i}{\cdot}~ \D_i
}
 is a predicate definition. \form{{\exists} \setvars{w_i}{\cdot} ~\D_i} (i $\in$ {\em 1}...n) is
 a branch of the disjunction.
%
In each branch, we require that variables which are not in formal parameters must
be existentially quantified. 


\begin{defn}[Root Parameter]\savespace

Given {\ud} predicate \code{P}: 
\form{
 \seppred{\code{P}}{\setvars{v}}  {\equiv}
   \bigvee^n_{i=1} ({\exists} \setvars{w_i}{\cdot}~ \D_i );
},
a parameter \form{r {\in} \setvars{v}} is a {\bf root} if for any {\em base}
 formula \form{\heap{\wedge}\pure} derived by unfolding the predicate instance \form{\seppred{\code{P}}{\setvars{v}}},
\form{r} is either a root of a points-to predicate (\form{r\pto\_} occurs in \form{\heap}) or \form{r{=}\nil}.
\end{defn}
We syntactically detect root parameters as follows.
A formal parameter
\form{r {\in} \setvars{v}} is a {\bf root} if any branch \form{\heap_i{\wedge}\pure_i},
 \form{i {\in} \{1...n\}}, one of the following
four conditions holds:
(i) \form{r} is a root variable of a points-to predicate (\form{r\pto\_} occurs in \form{\heap_i});
(ii) r equals to \form{\nil}: \form{\pure_i {\implies} r{=}\nil};
 (iii) r equals to another root parameter:
 \form{\pure_i {\implies} }\form{r{=}s}, where \form{s {\in} \setvars{v}};
 or (iv) r is a root parameter of another {\ud} predicate.
Without loss of generality, we will write \seppred{\code{P}}{r,\setvars{v}}
to indicate that \form{r} is a root parameter.
A predicate with multiple root parameters will be considered to transform into
multiple predicates with single root parameter in Sec. \ref{sec:explore}. 

\hide{
invariants: 
\form{\inv(\code{\seppred{p}{\setvars{v}}}) \imply \invo(\code{\seppred{p}{\setvars{v}}})}.

\begin{verbatim}
exact: {} /\ x=p /\ n=0
   \/ {x} /\ n>0

over: n>=0

under: {} /\ x=p /\ n=0

x=null |- x::lsegn<null, 1>
x=null |- ({} /\ x=p /\ 1=0) \/ ({x} /\ 1>0)
x=null |-                        ({x} /\ 1>0)  -> fail

\end{verbatim}}

\noindent{\em Unfolding. }
The function \form{\unfold(\D, j)} unfolds once
 the j$^{th}$ {\ud} predicate instance, i.e. \code{\seppred{\code{P^j}}{\setvars{t}^j}},
  of the formula
\form{\D}. The steps are formalized as follows:
\savespace\[\savespace
\begin{array}{c}
\seppred{\code{P^j}}{\setvars{v}}  {\equiv} \bigvee^n_{i=1} (\exists \setvars{w}_i \cdot ~\heap_i {\wedge} \pure_i)   \quad 
\fresh ~ \setvars{w}^{'}_i \quad  
 \rho_i {=}  [\setvars{w}^{'}_i / \setvars{w_i}] \quad
\heap'_i = \heap_i [\rho_i ] \quad \pure'_i = \pure_i [\rho_i ]\\
\rho_0 {=}  [\setvars{t}^i / \setvars{v}] \qquad
\heap''_i = \heap'_i [\rho_0] \quad \pure''_i = \pure'_i [\rho_0]\\ 
\hline
\unfold(\exists \setvars{w}_0 {\cdot}~  \seppred{\code{P^j}}{\setvars{t}^j} {\sep} \heap_0 {\wedge} \pure_0, \code{j}) \yields
\bigvee^n_{i=1} (\exists \setvars{w}_0 {\cup} \setvars{w}^{'}_i {\cdot}~ \heap_0 \sep \heap''_i {\wedge} \pure_0 {\wedge} \pure''_i)
\end{array}
\]
First, the function looks up the definition of \code{P},
refreshes the existential quantifiers.
Second, formal parameters are substituted by the actual parameters.
Finally, substituted definition is combined (and normalized) with residual formula
as in the RHS of \form{\yields}.
We will refer to
recursive predicate instances of unfolded heap formula \form{\heap_i} as
{\em descendant} predicate instances (of the j$^{th}$ {\ud} predicate instance).







\section{Lemma Mechanism} 

\label{sec:lemma}

\noindent {\bf Lemma Formalism.}\label{lem}
In general, lemmas in our system are formalized as:
\savespace\[\savespace
 \form{\code{lemma} ~ \code{id{:}}~ \forall \setvars{v} {\cdot}
\exists \setvars{w}_1{\cdot} \head_{l}{\sep}\D_l(\setvars{v}) ~{\rightarrow}~
\exists \setvars{w}_2{\cdot} \head_{r}{\sep}\D_r(\setvars{v})}.
\]
Our lemmas are used to relate two reachable heap regions starting from
 a same root pointer of the root predicates \form{\head_{l}} and \form{\head_{r}}.
A root predicate is a points-to or a {\ud} predicate.
Without loss of generality, we require that 
the root pointer of a root predicate is explicitly denoted
with the preserved name \form{\self}.
For frame inference,
LHS and RHS must capture the same heaps
(i.e.
\form{ \forall \setvars{v} {\cdot}\exists \setvars{w}_1{\cdot} \head_{l}{\sep}\D_l(\setvars{v}) 
~{\ent_L}~ \exists \setvars{w}_2{\cdot} \head_{r}{\sep}\D_r(\setvars{v}) {\yields} (\_, \textcolor{blue}{\emp}{\wedge}\pure)}.
We occasionally use \code{id} to indicate the lemma.

\hide{Compared with  \cite{Nguyen:CAV08}, our
 lemma mechanism 
has two 
 interesting extensions. 
First, for application of lemma with universal variables,
 while complex lemma \cite{Nguyen:CAV08}
is based on ``delayed guard'',
our universal lemma is based on eager guard instantiation.
Second, 
out of the focus of
\cite{Nguyen:CAV08},
our mechanism is empowered with {\sobd} 
which is essential to lemma synthesis.
Third, our mechanism is more expressive than the one in \cite{Nguyen:CAV08}.
As the mechanism in \cite{Nguyen:CAV08}
was designed for alternative unfoldings beyond
{\ud} 
predicates' definitions, root predicates of
the complex lemmas are restricted on 
{\ud} 
 predicates
 (e.g. see rule \code{L{-}LEFT{-}COMPLEX}, Sec. 4 in \cite{Nguyen:CAV08}).
Our mechanism is targeted on inductive reasoning, our root predicates
can be either {\ud} 
 or points-to
predicates. (Thus, our mechanism can also be used to support for
 alternative unfoldings, like \cite{Nguyen:CAV08}.)}



\noindent{\bf Lemma Application. }\label{lem.app}
 During proof search,
 proven lemmas
 are considered as external inference rules. 
We assume that free variables of a lemma are
 renamed to avoid clashing before the lemma is applied. The application of a lemma \code{id}
is formalized as follows.
\savespace\[\savespace
\frac{\begin{array}{c}\entrulen{LAPP}\\
 (\code{lemma~id~} \head_{l} {\sep}\heap_l{\wedge}G {\rightarrow} \D_r) {\in} \lemstore
 \quad 
$\entailL{\D_1}{\heap_l[\rho]~ {\wedge}~G[\rho]}{\lemstore}{\RA_1}{\constr_{r_1}} $\\
{\rho} ~{=}~  \code{match}(\head_{l},\head_{1})
\qquad {\entailL{\constr_{r_1}{\sep} \D_{r}[\rho]}{\D_2}{\lemstore}{\RA_2}{\constr_{R}}}
\end{array}}
{\entailL{\head_1{\sep}\D_1}{\D_2}{\lemstore}{\RA_1 {\wedge}\RA_2}{\constr_R}}%
\]
The lemma \code{id} is applied into the above entailment through three steps.
First, we match the predicate \form{\head_{1}} of the antecedent 
 with root predicate \form{\head_l}. 
(We note that \form{\head_l} must contain a \form{\self} pointer.)
\form{\head_l} and \form{\head_{1}} are matched and unified by a partial function \code{match}.
If the matching is successful,  \code{match} produces substitutions as follows:
\savespace\[\savespace
\begin{array}{lcl}
\code{match}(\form{\sepnode{{\self}}{c}{\setvars{t}}}),\form{\sepnode{{x}}{c}{\setvars{w}}}) &{=} & \form{[\setvars{w}/\setvars{t}]\circ [x/ \self]} \\
\code{match}(\form{\seppred{\code{P}}{\self{,}\setvars{t}}},\form{\seppred{\code{P}}{x{,}\setvars{w}}}) &{=} & \form{[\setvars{w}/\setvars{t}]\circ [x/ \self]}
\end{array}
\]
Second, we prove guard and identify {\em cut} of the antecedent in the first line.
Last, we combine the residue of the antecedent with RHS of the
lemma before continuously proving the consequent (of the entailment) in the second line.


\noindent{\bf Universal Lemma Application.}
The universal guard \form{\forall G} is equivalent to {\em infinite} conjunction \form{\bigwedge_{\rho}G[\rho]}.
\form{\forall G} makes the lemmas with universal guard more expressive. However,
it also prevents applying these lemmas into entailments with instantiated (existential) guards
as the second step of the above lemma application rule could not be established
 (i.e. \form{\exists~x~ \D \not\ent\forall~x~ \D}).
While mechanism in \cite{Nguyen:CAV08} was based on ``delayed guard'',
we now propose {\em guard instantiation},
 a sound solution for universal lemma application.
In order to apply a universal lemma into
an entailment with instantiated (existential) guard,
we present a technique that dynamically and
 intelligently instantiates universal guard (of the lemma)
before really deploying this lemma.
Concretely, our technique will substitute
universal guard by a {\em finite} set of its instances.
 The soundness of this technique
is based on the following theorem.

\begin{theorem}[Lemma Instantiation]
Let \form{\D_{ante}} and \form{\D_{conseq}} be antecedent and consequent of an entailment check.
Let \form{\lemstore} be a set of universal lemmas
and let \form{\lemstore'} be a set of instantiated lemmas obtained by substituting each 
\form{\forall \setvars{v} G(\setvars{v})} in a universal guard 
by a finite conjunction of its instantiations.
If \form{\D_{ante} ~{\ent_{\form{\lemstore}'}}~ \D_{conseq}} is valid
then
 so is \form{\D_{ante} ~{\ent_{\form{\lemstore}}}~ \D_{conseq}}. 
\end{theorem}
\savespace \savespace
\begin{proof}
Semantically, a guard with universal variables \form{\forall \setvars{v} G(\setvars{v})} implies
a conjunction of its instantiations.
As such universal guard \form{\forall \setvars{v} G(\setvars{v})} is on the left-hand side
of lemmas, a proof system with inference rules \form{\lemstore'} implies
the corresponding proof system with inference rules \form{\lemstore}.
Thus, if \form{\D_{ante} ~{\ent_{\form{\lemstore}'}}~ \D_{conseq}} is valid
then so is
 \form{\D_{ante} ~{\ent_{\form{\lemstore}}}~ \D_{conseq}}.
\qed
\end{proof}\savespace
Inspired by quantifier instantiation in Simplify \cite{Detlefs:2005:STP},
we symbolically select ``relevant'' instantiations over universal variables
of a guard by looking up substitutions over the universal variables
 such that
the instantiated lemma suffices to prove the validity of a given entailment.
The substitutions are selected through shape predicate matching \cite{Chin:SCP:12} (the
corresponding action used in \cite{Distefano:2008:OOPSLA} is subtracting.).
The application of lemmas with universal variables is formalized as:
\savespace
\[\savespace
\frac{\begin{array}{c}\entrulen{ENT-LAPP-\forall}\\
 (\code{lemma~id~}{\forall} \setvars{v} {\cdot}\head_l {\sep}\D_l{\wedge}G(\setvars{v}) {\rightarrow} \D_r) {\in} \lemstore
 \quad \rho{=}\rho_m{\circ} \code{ins}(\setvars{v}) \quad 
$\entailL{\D_1}{\D_l[\rho]}{\lemstore}{\RA_1}{\constr_{r_1}} $\\
{\rho_m} {=}  \code{match}(\head_l,\head_1)
\quad {\entailL{\constr_{r_1}{\sep} \D_{r}[\rho]}{\D_2}{\lemstore}{\RA_2}{\constr_{r_2}}}
\quad {\entailL{\constr_{r_2}}{G(\setvars{v})[\rho]}{\lemstore}{\RA_3}{\constr_{R}}}
\end{array}}
{\entailL{\head_{1} {\sep}\D_1}{\D_2}{\lemstore}{\RA_1 {\wedge}\RA_2{\wedge}\RA_3}{\constr_R}}%
\]
In the rule above,
after head matching, we rename universal variables and obtain the substitution
by the function {\bf \code{ins}}. This substitution helps to instantiate
quantified variables 
in
both
universal guard of the lemma and existential guard of the antecedent.

Symbolic ``relevant" instantiations are selected during the course
of the left entailment at line 2 
and are captured in the residue \form{\constr_{r_2}}.
Finally, ``relevant" instantiations are used to
 prove the instantiated guard of the right entailment at line 2.
%
%
We implement the selection of the symbolic
 ``relevant" instantiations via
by instantiation
  mechanism in \cite{Chin:SCP:12}.
We summarize the mechanism in App. \ref{entail.rules}.








\section{Automated Inductive Entailment Procedure}
\label{ent.lemsyn}



\begin{figure}[tb]
\savespace
\begin{center}
\savespace 
\begin{frameit}
\savespace \savespace
\savespace\[\savespace
\!\Deq{\entrulen{R{\sep}}\\
\FV(\D_{c_1}) \cap \FV(\D_{c_2}){=}\{\} \qquad
{\entailL{\D_a}{\D_{c_1}}{\lemstore}{\RA_1}{\constr_{f_1}}}
\qquad {\entailL{\constr_{f_1}}{\D_{c_2}}{\lemstore}{\RA_2}{\constr_{f_2}}}
}
{
{\entailL{\D_a}{\D_{c_1}{\sep}\D_{c_2}}{\lemstore}{\RA_1{\wedge}\RA_2}{\constr_{f_2}}}
}
\]
\savespace\[\savespace
\!\Deq{\entrulen{LSYN}\\
\code{lemsyn}(\head_l{\sep}\heap_a{\wedge}\pure_a{,} \head_c{\sep}\D_c){,}{\lemstore}) {\yields}
 (\code{lemma ~l{:}}
 \form{\forall \setvars{v} \forall \setvars{e} {\cdot} \head_a{\sep}\heap_a{\wedge}\FIL{\setvars{v}}{\pure_a} {\wedge}\code{P}(\setvars{v},\setvars{e}) {\rightarrow} {\head_c}{\sep}\D_c{\sep}\code{\UNK}(\setvars{v})}{,} \RA) 
 \\
 {\rho_m} ~{=}~  \code{match}(\head_{l},\head_{a}) 
\quad \rho_i{=}\code{ins}(\setvars{e}) \quad \rho{=}\rho_m{\circ} \rho_i
 \\
{\entailL{(\head_a {\sep} \heap_a{\wedge} \pure_a)[\rho_i]}{(\head_a{\sep}\heap_a{\wedge}\FIL{\setvars{v}}{\pure_a} {\wedge}\code{P}(\setvars{v},\setvars{e}))[\rho]}{\lemstore}{\true}{\emp{\wedge}\pure_{f_1}}}
}
{
{\entailL{\forall \setvars{v} {\exists} \setvars{e}{\cdot}\head_a {\sep} \heap_a{\wedge}\pure_a}{\head_c{\sep}\D_c}{\lemstore}{\RA}{\code{\UNK}(\setvars{v})[\rho]}}
}
\]
\[\savespace
\begin{array}{l}
\!\Deq{\entrulen{AU}\\
\form{\horn~{\equiv}~
\seppredF{\code{\UNK}}{r{,}\setvars{w},\setvars{z}\NI}
~{\imply}~
\head(r{,}\setvars{t}){\sep} \seppredF{\code{\UNK_f}}{\setvars{w},\setvars{t'},\setvars{z}\NI,r\NI} {\wedge}\FIL{\setvars{w}\cup\{r\}}{\pure_1}
}
\\
\setvars{t'}=\setvars{t}\setminus (\setvars{w} \cup \setvars{z}\cup \{r\}) \qquad
\entailSYN{\seppredF{\code{\UNK_f}}{\setvars{w},\setvars{t'},\setvars{z}\NI,r\NI} {\sep}\heap_{1}{\wedge}\pure_1}{\lemstore}
{\heap_2 {\wedge}\pure_2}{\Horn}{\D_f}
}
{\entailSYN{\seppredF{\code{\UNK}}{r,\setvars{w},\setvars{z}\NI}{\sep} \heap_1 {\wedge}\pure_1}{\lemstore}
{\head(r{,}\setvars{t}) {\sep} \heap_2
{\wedge}\pure_2}{\horn {\wedge} \Horn}{\D_f}}
\end{array}
\]
\[
\begin{array}{l}
\!\Deq{\entrulen{AF}\\
\form{\horn~{\equiv}~\head(r{,}\setvars{t}){\sep} \seppredF{\code{\UNK_f}}{\setvars{w},\setvars{t'},\setvars{z}\NI,r\NI} {\wedge}\FIL{\setvars{w}\cup\{r\}}{\pure_1}~{\imply}~
\seppredF{\UNK}{r{,}\setvars{w},\setvars{z}\NI}}
\\
\setvars{t'}=\setvars{t}\setminus (\setvars{w} \cup \setvars{z}\cup \{t\}) \qquad
\entailSYN{\heap_{1}~{\wedge}~\pure_1}{\lemstore}
{\seppredF{\code{\UNK_f}}{\setvars{w},\setvars{t'},\setvars{z}\NI,r\NI} {\sep}\heap_2 ~{\wedge}~\pure_2}{\Horn}{\D_f}
}
{\entailSYN{ \head(r{,}\setvars{t}) {\sep}\heap_1 {\wedge}\pure_1}{\lemstore}
{\seppredF{\code{\UNK}}{r,\setvars{w},\setvars{z}\NI}{\sep}\heap_2
{\wedge}\pure_2}{\horn {\wedge} \Horn}{\D_f}}
\end{array}
\]
\savespace\[\savespace
\begin{array}{l}
\!\Deq{\entrulen{AU-P}\\
\pure{\equiv}\forall (\FV(\pure_1) \cup \FV(\pure_2)\setminus \setvars{w}) \neg{\pure_1} {\vee} {\pure_2} \qquad
\pure {\neq}\false \qquad
\code{\horn~{\equiv}~  \seppredF{\code{\UNKP}}{\setvars{w}} ~{\imply}~
\pure}
}
{\entailSYN{ \heap_1 {\wedge} \seppredF{\code{\UNKP}}{\setvars{w}}{\wedge}\pure_1}{\lemstore}
{\emp {\wedge}\pure_2}{\horn}{\heap_1{\wedge}\pure_1}}
\end{array}
\]
\[
\begin{array}{l}
\!\Deq{\entrulen{AF-P}\\
\pure{=}\xpure({\heap_1}) {\wedge} \pure_1 \qquad
\pure ~{\implies}~\pure_2 
\qquad
\code{\horn~{\equiv}~ \FIL{\setvars{w}}{\pure}~{\imply}~
\seppredF{\UNKP}{\setvars{w}}}
}
{\entailSYN{ \heap_1 {\wedge}\pure_1}{\lemstore}
{\emp {\wedge} \seppredF{\code{\UNKP}}{\setvars{w}}
{\wedge}\pure_2}{\horn}{\heap_1{\wedge}\pure_1}}
\end{array}
\]
\label{fig.infer.rules}
\savespace 
\end{frameit}
\caption{Inference Rules for Lemma Synthesis with Predicate Inference.}
\end{center}
\savespace
\end{figure}

In Fig. \ref{fig.infer.rules}, we propose new six inference rules;
\form{\entrulen{R{*}}} is for cutting heaps;
 \form{\entrulen{LSYN}} is for
dynamically generating and proving auxiliary lemmas;
and the rest is for generating relational assumptions while checking entailment.
\form{\entrulen{R{*}}} rule, a revision (with frame inference)
  of the Monotonicity rule \cite{Ishtiaq:POPL01}, 
helps to generate smaller sub-goals, and then to generate
more reusable lemmas. 
In this rule,  $\FV$ returns free variables of a formula.
Two necessary conditions to apply the rule \form{\entrulen{LSYN}} are: (i)
 root predicates of the antecedent (LHS) and the consequent (RHS), \form{\head_a} and
 \form{\head_c}, must share a same root pointer;
(ii) \form{\head_r} must be a defined predicate instance i.e. \form{\seppredF{\code{P}}{x,\setvars{w}}}.
%
Here, \form{\FIL{\setvars{w}}{\pure}} is an auxiliary function that
existentially quantifies in {\pure} all free variables that are not in the set \form{\setvars{w}}.
We present the \code{lemsyn} procedure in the subsection \ref{sec.lemgen}.

In 
\form{\entrulen{AU}} and \form{\entrulen{AF}} rules,
\form{\head(r{,}\setvars{t})} is either \form{\sepnode{r}{c}{\setvars{t}}} or
known (defined) \form{\seppredF{\code{P}}{r{,}\setvars{t}}}, or unknown predicate  \form{\seppredF{\code{U'}}{r{,}\setvars{t}{,}\setvars{w}\NI}}.
We use \# notation in unknown predicates to guide abduction and proof search.
We only abduce on pointers without \form{\NI}-anotated.
\form{\seppredF{\code{\UNK_{f}}}{\setvars{w},\setvars{t'}}} is another
unknown predicate generated to capture downstream heaps.
After abduced pointers will be annotated with \form{\NI} to avoid
double abduction.
New unknown predicate \code{\UNK_{f}} is only generated if
 at least one parameter is not annotated with \form{\NI} (i.e. \form{\setvars{w}\cup\setvars{t'} {\neq} \emptyset}).
To avoid conflict between abduction rules
and other (unfolding, subtraction) during proof search, all root pointers
in a heap formula must be annotated with \form{\NI} in unknown predicates.
For examples,
in our system  while the formula \form{\sepnode{x}{\code{c_1}}{y}{\sep}\code{U_1}(x\textcolor{red}{\NI}{,}y)} is valid,
the formula \form{\sepnode{x}{\code{c_1}}{y}{\sep}\code{U_1}(x,y)} is invalid. For the check  \form{\sepnode{x}{\code{c_1}}{\nil} {\ent_{\lemstore}} \sepnode{x}{\code{c_1}}{y}{\sep}\code{U_1}(x{\NI}{,}y)}, our proof search will apply
subtraction the heap pointed by \form{x} rather than abdution.
We illustrate the \form{\entrulen{AF}} rule with the following example:
\savespace \savespace\begin{center}
\begin{small}
\def\defaultHypSeparation{\hskip 5pt}
\def\labelSpacing{1pt}
\AxiomC{ }
\LeftLabel{\scriptsize XPURE}
\UnaryInfC{$\begin{array}{l}
\form{\emp{\wedge}\true {\ent_{\lemstore}} \emp{\wedge}\true}
 {\yields} (\true {,}\emp{\wedge}\true)
\end{array}$}
\LeftLabel{\scriptsize AF}
\UnaryInfC{$\begin{array}{l}
\form{\sepnode{y}{\code{c_1}}{\nil}{\ent_{\lemstore}} \code{U_{2a}}(y{,}x\NI)}
 {\yields} (\horn_1{:~}\sepnode{y}{\code{c_1}}{\nil}{\wedge}\true \imply \code{U_2}(y{,}x\NI) {,}\emp{\wedge}\true)
\end{array}$}
\LeftLabel{\scriptsize AF}
\UnaryInfC{$\begin{array}{l}
\form{\sepnode{x}{\code{c_1}}{y}{\sep}\sepnode{y}{\code{c_1}}{\nil} {\ent_{\lemstore}} \code{U_2}(x{,}y)}
 {\yields} (\horn_1 {\wedge}~(\sepnode{x}{\code{c_1}}{y}{\sep} \code{U_{2a}}(y{,}x\NI) \imply \code{U_2}(x{,}y)){,} \emp{\wedge}\true)
\end{array}$}
\DisplayProof
\end{small}\end{center}
In the second application of \form{\entrulen{AF}} rule,
no new unknown predicate was introduced as there was no pointers without \form{\NI}-anotated.

\noindent \form{\entrulen{AU-P}} and \form{\entrulen{AF-P}} rules
abduce on pure and applied on consequent which has empty heap.
\form{\xpure} procedure soundly transforms a heap formula into a
pure formula \cite{Chin:SCP:12}.
\hide{\noindent{\bf Search Strategy. } 
Our proof search strategy applies inference rules
over  a {\ud} predicate instance following
 the decreasing order as:
matching, lemma application, lemma synthesis (and then lemma application)
and unfolding.
It is interesting to investigate a dynamic search strategy,
 like the strategy in Zeno \cite{Sonnex:2012:TACAS},
over these inference rules.}

\subsection{Lemma Synthesis - \code{lemsyn} Procedure}\label{sec.lemgen}
Given the entailment check
\form{\forall \setvars{v} {\exists} \setvars{e}{\cdot}\head_a {\sep} \heap_a{\wedge}\pure_a {\ent_{\lemstore}}\head_c{\sep}\D_r{\sep}\D_c},
our \code{lemsyn} procedure synthesizes the lemma
\form{\forall \setvars{v} \forall \setvars{e} {\cdot} \head_a{\sep}\heap_a{\wedge}\FIL{\setvars{v}}{\pure_a} {\wedge}\code{P}(\setvars{v},\setvars{e}) ~ {\rightarrow}~ {\head_c}{\sep}\D_r{\sep}\code{\UNK}(\setvars{v})}
 through three steps: conjecture construction,
lemma proving and predicate synthesis.

\noindent{\bf Conjecture Construction. }\label{sec.conjgen}
\hide{Generating a new conjecture is implemented by the function {\bf\code{syn}}
in three steps:
(i) constructing heap-only conjecture by syntactically finding reachable heaps of LHS and RHS from \form{x};
(ii) enriching the heap-only conjecture with (generated) unknown predicates;
and (iii) proving and synthesizing
 the conjecture using \code{lemprove} in Sec \ref{sec.lemprove}.
For (i), we define the function \code{rh} to collect
 reachable heaps as:
 \form{\code{rh}(r,\heap{\wedge}\pure){=}\sep \{\heap_i | (\heap_i \text{ occurs in } \heap)
 {\text{ and }} (\heap_i{\equiv}\sepnodeF{y}{c}{\_} \text{ or } \heap_i{\equiv} \seppredF{\code{P}}{y,\_})} and
\form{ y {\in} \code{reach}(r,\heap{\wedge}\pure)) \}}.
The function \form{\code{reach}(r,\D)} computes a set of pointers in \form{\D} which are reachable from \form{r}.
Assume \form{R{=}\code{reach}(r,\heap{\wedge}\pure)}, then the following are true:
\begin{itemize}
\item \form{r {\in} R}.
\item if \form{x {\in} R} and \form{\pure {\implies}x{=}y} then \form{y {\in} R}.
\item if \form{x {\in} R} and either \sepnodeF{x}{c}{\setvars{v}} or \seppredF{\code{P}}{x,\setvars{v}}
 occurs in \form{\heap} then 
for all \form{v_i} in \form{\setvars{v}}, \form{{v_i} {\in} R}.
Additionally, if {\form{\pure {\implies} x{\neq}r}} then we refer x as an {\em internal} variable.
\end{itemize}
We will refer a {\em leaf} variable
as a variable which is reachable from \form{r}
but is not internal.
For examples, reachable heaps from \form{x}
of LHS and RHS in the running example are:

\noindent \form{\code{rh}(x, \exists k{\cdot}\seppred{\code{lln}}{x,n} {\sep} \seppred{\code{lln}}{y,m} {\wedge} n{\geq}k {\wedge}k{\geq}0{\wedge}i{=}k{\wedge}j{=}n{-}k
)= \seppred{\code{lln}}{x,n}}; and

\noindent \form{\code{rh}(x, {\exists} p {\cdot}\seppred{\code{lsegn}}{x{,}p{,}i} {\sep}
\seppred{\code{lln}}{p,j}){=} \seppred{\code{lsegn}}{x{,}p{,}i} {\sep}
\seppred{\code{lln}}{p,j}}, respectively.
 And \form{n}, \form{i} and \form{j} are leaf variables. Heap-only conjecture is constructed as:
\[
\form{\code{lemma~sp~} 
\seppred{\code{lln}}{\self,n}
{\rightarrow}
\exists p {\cdot} \seppred{\code{lsegn}}{\self{,}p{,}i} {\sep}
 \seppred{\code{lln}}{p,j}}
\]

For (ii), in order to infer
guards 
 of conjecture besides the reachable heaps,
our system  generates one {\em unknown} predicate in LHS.
This unknown predicate takes all leaf variables
of LHS and only universal leaf variables of RHS as parameters.
These leaf variables 
are quantified as are their quantifiers in the original entailment.
With running example, our mechanism generates
an unknown predicate (e.g \form{P})
whose parameters (\form{n}, \form{i} and \form{j}) are universally quantified. Concretely,
 the following conjecture is generated:
\savespace\[\savespace
\form{\code{lemma~sp~} \forall n{,}i{,}j {\cdot} 
\seppred{\code{lln}}{\self,n} {\wedge} P(n{,}i{,}j)
\rightarrow
\exists p {\cdot} \seppred{\code{lsegn}}{\self{,}p{,}i} {\sep}
 \seppred{\code{lln}}{p,j}}
\]
For 
completeness, we always
consider to generate two-way lemmas. Thus,
  the reverse lemma of \code{sp} is also examined:
\savespace\[\savespace
\form{\code{lemma~jn~} \forall n{,}i{,}j {\cdot} 
\exists p {\cdot} \seppred{\code{lsegn}}{\self{,}p{,}i} {\sep}
 \seppred{\code{lln}}{p,j}
\rightarrow
\seppred{\code{lln}}{\self,n} {\wedge} P(n{,}i{,}j)
}
\]

For (iii), our system invokes the {\bf\code{lemprove}} procedure
to prove those generated conjectures and infer definitions for unknown predicates.}
%
%
At this step, \code{lemsyn} enriches the original check
with either pure unknown predicate \form{\code{P}(\setvars{v},\setvars{e})}
(for universal lemma) or shape unknown predicate \form{\code{\UNK}(\setvars{v})}
(for frame inference).
While the former is only added if there exist existential variables in LHS,
the latter is only added if the sets of universal variables over pointers of
LHS and RHS are not identical. We note that in the former, to prepare
for universal guard inference, we need
to existentially quantifies the pure formula;
in the latter, \form{\D_r} is a minimum closure
of connected heaps of \form{\head_c}. The unknown predicate is generated with
parameters that are union of free variables of LHS and RHS.
Among these, pointer-based parameters are annotated with \form{\NI} following the principle that
instantiation (and subtraction) are done before abduction.
The detail is as follows:
(i) all intersection variables of LHS and RHS are \form{\NI}-annotated;
(ii) roots pointers of RHS are \form{\NI}-annotated;
(iii) remaining pointers are not \form{\NI}-annotated.

\noindent{\bf Lemma Proving. }\label{sec.lemprove}
Our lemma proving, \code{lemprove} procedure,  is based on the principle of
cyclic proof \cite{Brotherston:CADE:11,Brotherston:APLAS:12}.
Two steps of this proof technique are:
 back-link form (i.e. linking current sequents to a historical sequent);
and {\em global trace condition} checking.
We implement these steps via lemma application.
%
%
The procedure \code{lemprove} is formalized as:
\[
\!\frac{\begin{array}{c}\entrulen{LEM-PROVE}\\
  {\bigvee^n_{i=1}} \D_i {\equiv} \form{\unfold(\D_l, \code{j})} \qquad  
{\entailL{\D_i}{\D_r}{\lemstore  {\cup} \code{\{link\}}}{\RA_i}{\textcolor{blue}{\emp}{\wedge}\pure_i}}
\end{array}}
{ (\code{lemma} ~ \code{link{:}}~ \form{{\D_l} ~{\rightarrow}~{\D_r}}, \lemstore) ~\form{{\yields}~\bigwedge^n_{i=1} \RA_i}} %
\]
To prove the conjecture \code{link}, \code{lemprove} looks
 up a $j^{th}$ \form{\ud} predicate instance in \form{\D_l} to
apply \form{\entrulen{LU}} rule.
For a successful proving, any disjunct \form{\D_i} obtained from the unfolding of
 \form{\D_l} must imply \form{\D_r} with {\em empty} heap
in the residue.
This unfolded predicate instance  is a progressing point.
Induction hypothesis is encoded by the lemma \code{link};
induction on such predicate instance is performed
as an application on this lemma into a descendant predicate instance of the unfolded predicate instance.
We denote such application is {\em cyclic} lemma application.
We note that the lemma synthesis may be nested; it means our system
would speculate additional lemmas while proving a lemma.

In cyclic term \cite{Brotherston:CADE:11,Brotherston:APLAS:12}, the initial lemma is a {\em bud},
entailment check which applied cyclic lemma application is 
{\em companion} of the bud above, and proof that are removed all cyclic lemma applications is a pre-proof.
A path in a pre-proof is a sequence of sequent occurrences (entailment checks)
derived by applying inference rules.
For soundness, a pre-proof must be a cyclic proof;
it must satisfy the global trace condition
 i.e. for every infinite path there is infinitely many progressing points.
%
We state the condition that a proof derived by our system is indeed a cyclic proof
in the following lemma.
\begin{lemma}[Soundness]
If all descendant predicate instances of the unfolded $j^{th}$ \form{\ud} predicate instance
have involved in a cyclic lemma application, then
\form{\bigwedge^n_{i=1} \RA_i {\wedge}\D_l ~{\models}~\D_r}.
\end{lemma}
\begin{proof}
A path in the proof is infinite if it includes the descendant predicate instances
(of the $j^{th}$ \form{\ud} predicate instances)
as
these predicate instances may unfolded infinitely.
If every the descendant predicate instances 
has involved in a cyclic lemma application,
it has infinitely progressing points and thus
 the global trace condition is satisfied.
Then our proof is a cyclic proof.
\qed
\end{proof}





\noindent{\bf Predicate Synthesis. }
If the conjecture \code{link} contains unknown predicates, we infer by {\sobd} a conjunction set of
relational constraints \form{\RA{=}\bigwedge^n_{i=1} \RA_i} (over the unknown predicates)
such that \form{\RA {\wedge} \D_l \models \D_r}.
After that we deploy those algorithms in \cite{Loc:CAV:2014,Popeea:ASIAN06} to solve
the set \form{\RA}
 and obtain definitions for the predicates.
Due to the nested lemma synthesis, a predicate is synthesis at the scope
it has been introduced.
After that its corresponding
 relational assumptions are canceled, i.e. not forwarded to
outer scope.

\paragraph{Two-way Lemmas.} For each lemma synthesized, we always
consider to generate its reverse lemma.
For each pair of such {\em two-way} lemmas,
one with unknown predicates will be inferred (and proven);
another is substituted with the newly-inferred predicates
prior to proven. For the former, 
we choose the conjecture with more case splits i.e. more {\ud} predicates
in the LHS. 
We hope that
proving such conjecture will  generate more subgoals
and thus more relational assumptions would be generated.
The more assumptions our system generates, the more
meaningful predicate definitions is synthesized.

\subsection{Motivating Example Revisit}\label{sec.lemprove.examp}
Our system started proving \code{l_3} by unfolding predicate
\form{\seppred{\code{ls1}}{a{,}p_1}} (like \form{\entrulen{LU}}) as follows.
 \savespace \begin{center}
\begin{small}
\def\defaultHypSeparation{\hskip 0pt}
\def\labelSpacing{0pt}
\AxiomC{$\begin{array}{l}
\form{\seppred{\code{ls2}}{a{,}p_2}{\sep}
 \sepnode{p_2}{\code{c_2}}{3,\nil}}\\
\form{
~~{\ent_{\{l_3\}}}
\seppred{\code{ls1}}{a{,}p_3}{\sep}\sepnode{p_3}{\code{c_2}}{v_1{,}p_4}{\sep}
\code{U}(\setvars{v_1}){\wedge}v_1{\neq}1
}
\end{array}$}
\AxiomC{$\begin{array}{l}
\form{\sepnode{a}{\code{c_2}}{1,a_1}{\sep}\underline{\seppred{\code{ls1}}{a_1{,}p_1}} {\sep}\seppred{\code{ls2}}{p_1{,}p_2}{\sep} \\
~~ \sepnode{p_2}{\code{c_2}}{3,\nil}} 
\form{{~\ent_{\{l_3\}}}~
\D_c
}
\end{array}$}
\BinaryInfC{$\begin{array}{l}
\form{\underline{\seppred{\code{ls1}}{a{,}p_1}} {\sep}\seppred{\code{ls2}}{p_1{,}p_2}{\sep}
 \sepnode{p_2}{\code{c_2}}{3,\nil}}
\form{{\ent_{\emptyset}}
\seppred{\code{ls1}}{a{,}p_3}{\sep}\sepnode{p_3}{\code{c_2}}{v_1{,}p_4}{\sep}\code{U}(\setvars{v}){\wedge}v_1{\neq}1
}
\end{array}$}
\DisplayProof
\end{small}\end{center}\saveone
whereas
\form{\D_c{\equiv}\seppred{\code{ls1}}{a{,}p_3}{\sep}\sepnode{p_3}{\code{c_2}}{v_1{,}p_4}{\sep}\code{U}(a\NI{,}p_1{,}p_2{,}p_3\NI{,}p_4{,}v_1){\wedge}v_1{\neq}1}, and

\noindent \form{\setvars{v}{\equiv}\{a\NI{,}p_1{,}p_2{,}p_3\NI{,}p_4{,}v_1}\},
 \form{\setvars{v_1}{\equiv}\{a\NI{,}a{,}p_2{,}p_3\NI{,}p_4{,}v_1}\}.
(In all tree derivations below, we discard the \form{\entrulen{XPURE}} rule on top for simplicity.)

\noindent{\bf Base Case. } Proof of the left subgoal was derived as:
 \savespace \savespace \begin{center}
\begin{small} \savespace
\def\defaultHypSeparation{\hskip .1in}
\def\labelSpacing{1pt}
\AxiomC{$\begin{array}{l}
 \form{\emp{\wedge}\true}{\ent_{\{l_3;l_4\}}} \form{\emp{\wedge}\true} \yields (\true,\emp{\wedge}\true)
\end{array}$}
\LeftLabel{\scriptsize LSYN}\RightLabel{\scriptsize \RA}
\UnaryInfC{$\begin{array}{l}
\form{\seppred{\code{ls2}}{a{,}p_2}{\sep}
 \sepnode{p_2}{\code{c_2}}{3,\nil}}
\form{{\ent_{\{l_3\}}}\sepnode{a}{\code{c_2}}{v_1{,}p_4}{\sep}\code{U}(\setvars{v_2}){\wedge}v_1{\neq}1}
\end{array}$}
\LeftLabel{\scriptsize RU}
\UnaryInfC{$\begin{array}{l}
\form{\seppred{\code{ls2}}{a{,}p_2}{\sep}
 \sepnode{p_2}{\code{c_2}}{3,\nil}}
\form{{\ent_{\{l_3\}}}\seppred{\code{ls1}}{a{,}p_3}{\sep}\sepnode{p_3}{\code{c_2}}{v_1{,}p_4}{\sep}\code{U}(\setvars{v_1}){\wedge}v_1{\neq}1{\yields}(\RA_1{,}\emp{\wedge}..)}
\end{array}$}
\DisplayProof
\end{small}\end{center}
whereas \form{\setvars{v_2}{\equiv}\{a\NI{,}a{,}p_2{,}a\NI{,}p_4{,}v_1}\}, and
the nested conjecture \code{l_4} was constructed as:
\savespace \saveone\[\savespace \saveone
\code{lemma~l_4:}\form{\seppred{\code{ls2}}{a{,}p_2}{\sep}
 \sepnode{p_2}{\code{c_2}}{3,\nil}
{\rightarrow}\sepnode{a}{\code{c_2}}{v_1{,}p_4}{\sep}\code{U}(\setvars{v_2}){\wedge}v_1{\neq}1 }
\]
Proof of the conjecture \code{l_4} was derived as in Fig. \ref{fig.prove.l4},
\begin{figure}[tb]
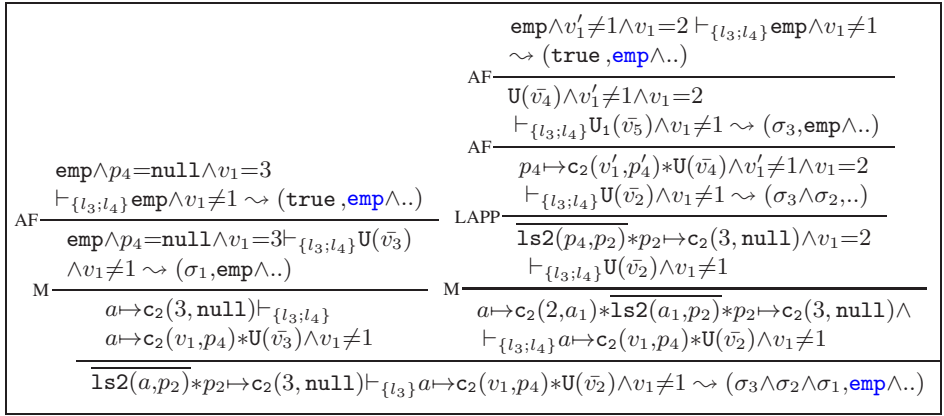

\savespace \savespace \begin{center}
\begin{small}\savespace \saveone
\begin{frameit} \savespace \saveone
\savespace \savespace\[\savespace
\def\defaultHypSeparation{\hskip 2pt}
\def\labelSpacing{1pt}
\AxiomC{$\begin{array}{l}
\form{\emp{\wedge}p_4{=}\nil{\wedge}v_1{=}3}\\
\form{{\ent_{\{l_3;l_4\}}}\emp{\wedge}v_1{\neq}1 
\yields (\true{,}\textcolor{blue}{\emp}{\wedge}..)}
\end{array}$}
\LeftLabel{\scriptsize AF}
\UnaryInfC{$\begin{array}{l}
\form{\emp{\wedge} p_4{=}\nil{\wedge}v_1{=}3\ent_{\{l_3;l_4\}}\code{U}(\setvars{v_3})}\\
\form{ {\wedge}v_1{\neq}1
\yields (\horn_1{,}{\emp}{\wedge}..)}
\end{array}$}
\LeftLabel{\scriptsize M}
\UnaryInfC{$\begin{array}{l}
\form{ \sepnode{a}{\code{c_2}}{3,\nil}{\ent_{\{l_3;l_4\}}}}\\
\form{\sepnode{a}{\code{c_2}}{v_1{,}p_4}{\sep}\code{\UNK}(\setvars{v_3}){\wedge}v_1{\neq}1}
\end{array}$}
\AxiomC{$\begin{array}{l}
\form{\emp{\wedge}v_1'{\neq}1 {\wedge}v_1{=}2}
~\form{{\ent_{\{l_3;l_4\}}}\emp{\wedge}v_1{\neq}1 \\
\yields (\true{,}\textcolor{blue}{\emp}{\wedge}..)}
\end{array}$}
\LeftLabel{\scriptsize AF}
\UnaryInfC{$\begin{array}{l}
\form{\code{U}(\setvars{v_4}){\wedge}v_1'{\neq}1 {\wedge}v_1{=}2}\\
~\form{{\ent_{\{l_3;l_4\}}}\code{U_1}(\setvars{v_5}){\wedge}v_1{\neq}1
\yields (\horn_3{,}\emp{\wedge}..)}
\end{array}$}
\LeftLabel{\scriptsize AF}
\UnaryInfC{$\begin{array}{l}
\form{\sepnode{p_4}{\code{c_2}}{v_1'{,}p_4'}{\sep}\code{U}(\setvars{v_4}){\wedge}v_1'{\neq}1 {\wedge}v_1{=}2 }\\
~\form{{\ent_{\{l_3;l_4\}}}\code{U}(\setvars{v_2}){\wedge}v_1{\neq}1
\yields (\horn_3{\wedge}\horn_2{,}..)}
\end{array}$}
\LeftLabel{\scriptsize LAPP}
\UnaryInfC{$\begin{array}{l}
\form{\overline{\seppred{\code{ls2}}{p_4{,}p_2}}{\sep}
 \sepnode{p_2}{\code{c_2}}{3,\nil}{\wedge}v_1{=}2}\\
~\form{~{\ent_{\{l_3;l_4\}}}\code{U}(\setvars{v_2}){\wedge}v_1{\neq}1}
\end{array}$}
\LeftLabel{\scriptsize M}
\UnaryInfC{$\begin{array}{l}
\form{\sepnode{a}{\code{c_2}}{2{,}a_1}{\sep}\overline{\seppred{\code{ls2}}{a_1{,}p_2}}{\sep}
 \sepnode{p_2}{\code{c_2}}{3,\nil}{\wedge}}\\
~\form{{\ent_{\{l_3;l_4\}}}\sepnode{a}{\code{c_2}}{v_1{,}p_4}{\sep}
\code{U}(\setvars{v_2}){\wedge}v_1{\neq}1}
\end{array}$}
\BinaryInfC{$\begin{array}{l}
\form{\overline{\seppred{\code{ls2}}{a{,}p_2}}{\sep}
 \sepnode{p_2}{\code{c_2}}{3,\nil}}
\form{{\ent_{\{l_3\}}}\sepnode{a}{\code{c_2}}{v_1{,}p_4}{\sep}\code{U}(\setvars{v_2}){\wedge}v_1{\neq}1}\yields (\horn_3{\wedge}\horn_2{\wedge}\horn_1{,}\textcolor{blue}{\emp}{\wedge}..)
\end{array}$}
\DisplayProof\]
\end{frameit}\saveone
\caption{Derivation tree for proving of lemma \code{l_4}.}
\label{fig.prove.l4}
\end{small} 
\end{center}
\end{figure}
whereas 
\form{(\setvars{v_3}){\equiv}(a\NI{,}a{,}a{,}a\NI{,}p_4{,}v_1}),
\form{(\setvars{v_4}){\equiv}(a\NI{,}a{,}p_2{,}a\NI{,}p_4'{,}v_1'}),
\form{(\setvars{v_5}){\equiv}(a\NI{,}a{,}p_2{,}a\NI{,}p_4\NI{,}v_1{,}p_4'{,}v_1'}).
Inferred assumptions are: 
\[
\begin{array}{l}
\form{\horn_1{:}}~ \form{p_4{=}\nil {\wedge}v_1{=}3}
 \form{~{\imply}~{\code{U}(a{,}a{,}a{,}a{,}p_4{,}v_1)}}\\
\form{\horn_2{:}}~ \form{\sepnode{p_4}{\code{c_2}}{v_1'{,}p_4'}
{\sep}\code{U_1}(a{,}a{,}p_2{,}a{,}p_4{,}v_1{,}p_4'{,}v_1')
{\wedge}
 {v_1'{\neq}1}{\wedge}v_1{=}2}
 \form{~{\imply}~{\code{U}(a{,}a{,}p_2{,}a{,}p_4{,}v_1)}}\\
\form{\horn_3{:}}~ \form{\code{U}(a{,}a{,}p_2{,}a{,}p_4'{,}v_1')
{\wedge}
 {v_1'{\neq}1}{\wedge}v_1{=}2}
 \form{~{\imply}~ \code{U_1}(a{,}a{,}p_2{,}a{,}p_4{,}v_1{,}p_4'{,}v_1')} \\
\end{array}
\]
Since \code{U_1} was introduced at this scope, before return to outer scope,
\code{U_1} has been synthesized as
\form{\defpred~ \code{\code{U_1}}(a{,}a{,}p_2{,}a{,}p_4{,}v_1{,}\self{,}v_1'){\equiv}
\exists p_4{\cdot}\code{U}(a{,}a{,}p_2{,}a{,}\self{,}v_1'){\wedge}
 {v_1'{\neq}1}{\wedge}v_1{=}2}.
The set assumptions forwarded to outer scope is: \form{\RA_1{\equiv}\horn_1{\wedge}\horn_2}.

\noindent{\bf Induction Case. } Proof of the right subgoal was derived as:
\savespace \savespace \saveone \begin{center}
\begin{small}
\def\defaultHypSeparation{\hskip .1in}
\def\labelSpacing{1pt}
\AxiomC{$\begin{array}{l}
\form{\emp{\wedge}v_1{\neq}1}
~\form{{\ent_{\{l_3\}}}~\emp{\wedge}v_1{\neq}1\yields (\horn_4{,}\textcolor{blue}{\emp}{\wedge}..)}
\end{array}$}
\LeftLabel{\scriptsize AF}
\UnaryInfC{$\begin{array}{l}
\form{\code{U}(\setvars{v_7}){\wedge}v_1{\neq}1}
~\form{{\ent_{\{l_3\}}}~\code{U}(\setvars{v_1}){\wedge}v_1{\neq}1\yields (\horn_4,\emp{\wedge}..)}
\end{array}$}
\LeftLabel{\scriptsize M}
\UnaryInfC{$\begin{array}{l}
\form{\seppred{\code{ls1}}{a_1{,}p_3'}{\sep}\sepnode{p_3'}{\code{c_2}}{v_1'{,}p_4'}{\sep}\code{U}(\setvars{v_6}){\wedge}v_1'{\neq}1} 
~\form{{\ent_{\{l_3\}}}\seppred{\code{ls1}}{a_1{,}p_3}{\sep}\sepnode{p_3}{\code{c_2}}{v_1{,}p_4}{\sep}\code{U}(\setvars{v}){\wedge}v_1{\neq}1}
\end{array}$}
\LeftLabel{\scriptsize LAPP}
\UnaryInfC{$\begin{array}{l}
\form{\underline{\seppred{\code{ls1}}{a_1{,}p_1}}
 {\sep}\seppred{\code{ls2}}{p_1{,}p_2}{\sep}
 \sepnode{p_2}{\code{c_2}}{3,\nil}} 
~\form{{\ent_{\{l_3\}}}\seppred{\code{ls1}}{a_1{,}p_3}{\sep}\sepnode{p_3}{\code{c_2}}{v_1{,}p_4}{\sep}\code{U}(\setvars{v}){\wedge}v_1{\neq}1}
\end{array}$}
\LeftLabel{\scriptsize M}
\UnaryInfC{$\begin{array}{l}
\form{\sepnode{a}{\code{c_2}}{1,a_1}{\sep}\underline{\seppred{\code{ls1}}{a_1{,}p_1}}
 {\sep}\seppred{\code{ls2}}{p_1{,}p_2}{\sep}
 \sepnode{p_2}{\code{c_2}}{3,\nil}} \\
~~\quad\form{{\ent_{\{l_3\}}}\sepnode{a}{\code{c_2}}{1,a_1}{\sep}\seppred{\code{ls1}}{a_1{,}p_3}{\sep}\sepnode{p_3}{\code{c_2}}{v_1{,}p_4}{\sep}\code{U}(\setvars{v}){\wedge}v_1{\neq}1}
\end{array}$}
\LeftLabel{\scriptsize RU}
\UnaryInfC{$\begin{array}{l}
\form{\sepnode{a}{\code{c_2}}{1,a_1}{\sep}\underline{\seppred{\code{ls1}}{a_1{,}p_1}} {\sep}\seppred{\code{ls2}}{p_1{,}p_2}{\sep}
 \sepnode{p_2}{\code{c_2}}{3,\nil}}
~\form{{\ent_{\{l_3\}}}~
\D_c
{\yields} (\horn_4{,}\textcolor{blue}{\emp}{\wedge}..)}
\end{array}$}
\DisplayProof\savespace
\end{small}\end{center} \saveone
whereas \form{(\setvars{v_6}){\equiv}(a_1\NI{,}p_1{,}p_2{,}p_3'\NI{,}p_4'{,}v_1')},
\form{(\setvars{v_7}){\equiv}(a_1\NI{,}p_1{,}p_2{,}p_3\NI{,}p_4{,}v_1)}, and
the inferred assumption is:
\form{\horn_4{:}}~ \form{\code{U}(a_1{,}p_1{,}p_2{,}p_3{,}p_4{,}v_1){\wedge}v_1{\neq}1}~
 \form{~{\imply}~{\code{U}(a{,}p_1{,}p_2{,}p_3{,}p_4{,}v_1)}}\\
Now, the predicate \code{U} is synthesized from
the set of assumptions \form{\horn_1{\wedge}\horn_2{\wedge}\horn_4}.

\subsection{Soundness}\label{sec.entail.sound}
\saveone\begin{lemma}
The rules \form{\entrulen{R{\sep}}} and 
\form{\entrulen{LSYN}} in Fig. \ref{fig.infer.rules} preserve soundness.
\end{lemma}\saveone
Our \form{\entrulen{R{\sep}}} rule is derived
from the Monotonicity rule showing that spatial conjunction is monotone
with respect to implication \cite{Ishtiaq:POPL01}:
\savespace \saveone\[\savespace \saveone
\!\Deq{
{\D_{a_1}}{\models}{\D_{c_1}}
\qquad {{\D_{a_2}}{\models}{\D_{c_2}}}
}
{
{\D_{a_1}{\sep}\D_{a_2}} {\models} {\D_{c_1}{\sep}\D_{c_2}}
}
\]
The soundness of \form{\entrulen{LSYN}} is derived from the meaning of frame,
i.e. \form{\D_a {\ent} \D_c {\yields}\D_{frame}} holds iff \form{\D_a{\models}\D_c{\sep}\D_{frame}},
and the design of our lemma mechanism,
 i.e.  \form{\code{lemma ~l{:}} \D_l {\rightarrow} \D_r } is valid iff
 \form{\D_l {\ent} \D_r {\yields}\emp} holds.


\noindent{\bf Soundness of abduction rules.} Since we only apply
proven lemmas, it is sound to assume that lemma store is empty (no user-supplied lemmas) and discard this lemma store in our soundness proofs.
 We introduce the notation
\form{\assumpset(\Gamma)} to denote a set of predicate definitions
\form{\Gamma{=}\{\form{\UNK_1(\bar{v}_1){\implyeq}\constr_1},..\form{\UNK_n(\bar{v}_n){\implyeq}\constr_n}\}}
satisfying the set of assumptions \sm{\assumpset}.
That is, for all assumptions \form{{\D_l\imply\D_r}\in\assumpset}, (i)
\form{\Gamma} contains a predicate definition for each unknown predicate appearing in
\form{\D_l} and \form{\D_r};
(ii) by interpreting all unknown predicates according to \form{\Gamma}, then it is
provable that \form{\D_l} implies \form{\D_r}, written as
  \form{\code{\Gamma}~{:}~\D_l\vdash\D_r}.
\saveone
\begin{lemma}\label{lem.bi}
Given the entailment judgement
\form{ \entailSA{\setvars{U}}{\D_{a}} {\D_{c}}
{\RA}{\D_{f}} },
if there exists \form{{\Gamma}}  such that \form{\assumpset(\Gamma)},
then the entailment \form{{\Gamma}: \D_{a}\vdash
\D_{c} \sep \D_{f}} holds.
\end{lemma} \saveone
  Abduction soundness requires that if all the relational assumptions
generated are satisfiable, then the entailment is valid.

\section{Theorem Exploration}
\label{sec:explore}
We present a mechanism to explore relations for
{\ud} predicates by using lemma synthesis.
 The mechanism is applied for
some sets
 of either statically user-supplied or dynamically analyser-synthesized predicates.
The key idea is that instead of requiring designers of proof systems
to write lemmas for a specific {\ud} predicate (like in \cite{Berdine:APLAS05}),
we provide for them a mechanism to design lemmas for a
general class
of {\ud} predicates.
And based on the design,
 our system will automatically generate specific and on-demand lemmas
for a specific {\ud} predicate.
In this section, we demonstrate this mechanism
through three such classes. Concretely,
our system processes each {\ud} predicate in
four steps. First, it syntactically classifies the predicate into a predefined class.
Second, it follows 
 structure
 of the class 
 to generate {\em heap-only} conjectures (with quantifiers).
Third, it enriches the heap-only conjectures with unknown predicates for
 expressive constraint inference.
Last,  it invokes the \code{lemprove} procedure to prove these conjectures,
infer definitions for the unknown predicates
and synthesize the lemmas.
Especially, whenever a universal lemma \form{L {\rightarrow} R} is proven,
its reverse (the lemma \form{R {\rightarrow} L}) is also examined.
Proven lemmas will be applied to enhance upcoming inductive proofs.

In the next subsection and App. \ref{app.pred.trans},
 we present theorem exploration in three classes of {\ud} predicates.
In each class, 
we present step 1  
and step 2;
steps 3 and 4 are identical to 
 the 
\code{lemsyn} procedure in Sec. \ref{sec.lemgen}.
\savespace
\subsection{Generating Equivalence Lemmas}\savespace
\noindent{\bf Step 1.} Intuitively, given a set \code{S} of {\ud} predicates
and another {\ud} predicate \code{P} (which is not in \code{S}),
we look up all predicates in \code{S} which are equivalent to \code{P}.
This exploration is applied to any new (either supplied or synthesized)
{\ud} predicate \code{P}.

\noindent{\bf Step 2.} Heap-only conjecture
 to explore equivalent relation of two predicates
(e.g. \form{\seppred{\code{P}}{x,\setvars{v}}} and \form{\seppred{\code{Q}}{x,\setvars{w}}})
is generated as: \form{
{\exists} \setvars{t_1}{\cdot} \seppred{\code{P}}{\self,\setvars{v}} {\rightarrow}
{\exists}\setvars{t_2}{\cdot} \seppred{\code{Q}}{\self,\setvars{w}}}, whereas
 \form{\setvars{t_1}{=}\setvars{v}{\setminus}\setvars{w}} and
\form{\setvars{t_2}{=}\setvars{w}{\setminus}\setvars{v}}. 
The shared root parameter \form{x} has been identified by examining
 all permutations of root parameters
 of the two predicates.
For example, 
with \code{lln} and \code{lsegn} in Sec. \ref{sec:motivate},
our system examines conjecture:
\form{\code{lemma~eq_l~}{\exists}p{\cdot}\seppred{\code{lsegn}}{\self{,}p{,}m}{\rightarrow}\seppred{\code{lln}}{\self{,}n}}.

\noindent At step 3, the unknown predicate \code{U} is added to infer constraints
over {\em leaf} variables \form{p}, \form{m}, and \form{n} as:
\form{\code{lemma~eq_l~} {\exists}p{\cdot}\seppred{\code{lsegn}}{\self{,}p{,}m}{\wedge}\seppred{\code{U_1}}{p{,}m{,}n}{\rightarrow} \seppred{\code{lln}}{\self{,}n} }.
At step 4, the conjecture \code{eq_l} is proven and a definition of \code{U_1} is inferred
as:
\form{\seppred{\code{U_1}}{p{,}m{,}n}{\equiv}p{=}\nil {\wedge}m{=}n}.
For the equivalence, our system also generates and proves
the reverse lemma of \code{~eq_l~}
as: \form{\code{lemma~eq_r~} \seppred{\code{lln}}{\self{,}n} {\rightarrow}{\exists}p{\cdot} \seppred{\code{lsegn}}{\self{,}p{,}m}{\wedge}p{=}\nil {\wedge}m{=}n}.

This technique can be applied
 to match a newly-inferred definition synthesized by
shape analyses (i.e. \cite{Brotherston-Gorogiannis:SAS:14,Loc:CAV:2014})
with existing predicates of a supplied library of predefined predicates.
For specification inference, we eagerly substitute a newly-inferred predicate
in specifications
by its equivalent-matching predicate from the library.
This makes inferred specifications  more
 understandable. Furthermore this also helps to avoid
 induction proving on proof obligations
generated from these specifications.




\savespace
\section{Implementation and Experiments}\label{sec.exp}
\savespace \saveone

\begin{table}[tb]
\begin{center} \savespace
\savespace
\begin{tabular}[t]{| c|c | c | c | c|c|}
\hline
   Ent. & Proven & {Cyclic$_{SL}$} & {\toolname} & \#syn  \\
\hline
1 & \form{
 \seppred{\code{lseg}}{x{,}t}  {\sep}
 \seppred{\code{lseg}}{t{,}\nil}}  ~{\ent}{$_\emptyset$}~ \form{\seppred{\code{lseg}}{x{,}\nil}}& 0.03  & 0.06 & 1 \\ 
2 & \form{
 \seppred{\code{lseg}}{x{,}t} {\sep} \sepnode{\code{t}}{c_1}{y} {\sep}
 \seppred{\code{lseg}}{y{,}\nil}} ~{\ent}{$_\emptyset$}~ \form{\seppred{\code{lseg}}{x{,}\nil}}   &  0.03  & 0.08 & 1\\ 
3 &   \form{ 
 \seppred{\code{lseg}}{x{,}t}  {\sep}
 \seppred{\code{lseg}}{t{,}y}} {\sep} \sepnode{\code{y}}{c_1}{\nil}
  ~{\ent}{$_\emptyset$}~ \form{\seppred{\code{lseg}}{x{,}\nil}} & 0.03  & 0.11 & 2 \\ 
4 & \form{
 \seppred{\code{lseg}}{x{,}t}  {\sep}
 \seppred{\code{lseg}}{t{,}y} {\sep} \seppred{\code{bt}}{y}{\wedge}y{\neq}\nil}
 {\ent}{$_\emptyset$} \form{\seppred{\code{lseg}}{x{,}y}} {\sep} \seppred{\code{bt}}{y} & TO & 0.21 & 2 \\ 
5 & \form{
\seppred{\code{lseg}}{x{,}t}  {\sep}
 \seppred{\code{lseg}}{t{,}y} {\sep} \seppred{\code{lseg}}{y{,}z}{\wedge}y{\neq}z}
 {\ent}{$_\emptyset$} \form{\seppred{\code{lseg}}{x{,}y}} {\sep} \seppred{\code{lseg}}{y{,}z} & 3.00 & 0.57 & 1 \\
\hline
6 & \form{ 
 \sepnode{\code{x}}{c_1}{y} {\sep} \seppred{\code{rlseg}}{y{,}z}}
 ~ {\ent}{$_\emptyset$}~ \form{\seppred{\code{rlseg}}{x{,}z}} & 0.02  & 0.10 & 1 \\
 7 & \form{ 
 \seppred{\code{nlseg}}{x{,}z}{\sep}\sepnode{\code{z}}{c_1}{y}}
  ~ {\ent}{$_\emptyset$}~ \form{\seppred{\code{nlseg}}{x{,}y}}  & 0.02  & 0.04 & 1 \\ 
8 & \form{
 \seppred{\code{nlseg}}{x{,}z}{\sep} \seppred{\code{nlseg}}{z{,}y}}
 ~ {\ent}{$_\emptyset$}~ \form{\seppred{\code{nlseg}}{x{,}y}}  &  0.03 & 0.06 & 1 \\
9 & \form{
\seppred{\code{glseg}}{x{,}z}{\sep} \sepnode{\code{z}}{c_1}{y} }
  ~ {\ent}{$_\emptyset$}~ \form{\seppred{\code{glseg}}{x{,}y}}  &  0.02 & 0.04 & 1 \\ 
10 & \form{
\seppred{\code{glseg}}{x{,}z}{\sep} \seppred{\code{glseg}}{z{,}y}}
 ~ {\ent}{$_\emptyset$}~ \form{\seppred{\code{glseg}}{x{,}y}}  &  0.02 & 0.04 & 1 \\
11 & \form{ \seppred{\code{dlseg}}{u{,}v{,}x{,}y}}
 ~ {\ent}{$_\emptyset$}~ \form{\seppred{\code{glseg_2}}{u{,}v}}  &  0.07 & 0.04 & 1 \\ 
12 & \form{ \seppred{\code{dlseg}}{w{,}v{,}x{,}z}{\sep}\seppred{\code{dlseg}}{u{,}w{,}z{,}y}}
 ~ {\ent}{$_\emptyset$}~ \form{\seppred{\code{dlseg}}{u{,}v{,}x{,}y}} & 0.04  & 0.11 & 1 \\
13 & \form{
\seppred{\code{listo}}{x{,}z}{\sep} \seppred{\code{listo}}{z{,}\nil}}
 ~ {\ent}{$_\emptyset$}~ \form{\seppred{\code{liste}}{x{,}\nil}}   & 0.06  & 0.06 & 2 \\
14 & \form{
 \seppred{\code{liste}}{x{,}z}{\sep} \seppred{\code{liste}}{z{,}\nil}}
 ~ {\ent}{$_\emptyset$}~ \form{\seppred{\code{liste}}{x{,}\nil}}   & 0.18   & 0.12 & 3 \\
15 & \form{
 \seppred{\code{listo}}{x{,}z}{\sep} \seppred{\code{liste}}{z{,}y}}
  ~ {\ent}{$_\emptyset$}~ \form{\seppred{\code{listo}}{x{,}y}}   &  4.33  & 0.88 & 3 \\ 
16 & \form{
\seppred{\code{binPath}}{x{,}z}{\sep} \seppred{\code{binPath}}{z{,}y}}
 ~ {\ent}{$_\emptyset$}~ \form{\seppred{\code{binPath}}{x{,}y}}   & 0.03 & 0.06 & 1 \\
17 & \form{\seppred{\code{binPath}}{x{,}y}}
 ~ {\ent}{$_\emptyset$}~ \form{\seppred{\code{binTreeSeg}}{x{,}y}}   &  0.12 &  0.08 & 1 \\
18 & \form{
 \seppred{\code{binTreeSeg}}{x{,}z}{\sep}\seppred{\code{binTreeSeg}}{z{,}y}}
 ~ {\ent}{$_\emptyset$}~ \form{\seppred{\code{binTreeSeg}}{x{,}y}}   & 0.20  & 0.66 & 1 \\
19 & \form{
 \seppred{\code{binTreeSeg}}{x{,}y}{\sep}\seppred{\code{binTree}}{y}}
 ~ {\ent}{$_\emptyset$}~ \form{\seppred{\code{binTree}}{x}}   & 0.06  & 0.03 & 1 \\
\hline
20 & \form{ \seppred{\code{sortll}}{x{,}min} }
 ~ {\ent}{$_\emptyset$}~ \form{\seppred{\code{ll}}{x}}   & X  & 0.05 & 1 \\
21 & \form{ \seppred{\code{sortlln}}{x{,}min{,}size} }
 ~ {\ent}{$_\emptyset$}~ \form{\seppred{\code{lln}}{x{,}size}}  & X  & 0.12 & 1 \\
22 & \form{ \seppred{\code{sortlln}}{x{,}min{,}size} }
 ~ {\ent}{$_\emptyset$}~ \form{\seppred{\code{sortll}}{x{,}min}}  & X  & 0.08 & 1 \\
23 & \form{ \seppred{\code{lsegn}}{x{,}y{,}sz_1} {\sep}\seppred{\code{lsegn}}{x{,}z{,}sz_2}}
 ~ {\ent}{$_\emptyset$}~ \form{\seppred{\code{lsegn}}{x{,}z{,}sz_1{+}sz_2}}  &  X   & 0.12 & 1 \\
24 & \form{ \seppred{\code{lsegn}}{x{,}y{,}size_1} {\sep}\seppred{\code{lsn}}{x{,}size_2}}
 ~ {\ent}{$_\emptyset$}~ \form{\seppred{\code{lsn}}{x{,}size_1{+}size_2}}  & X  & 0.10 & 1 \\
25 & \form{ \seppred{\code{lseg}}{x{,}tl} {\sep} \sepnode{\code{tl}}{c_1}{y} {\sep}
\seppred{\code{lseg1}}{y{,}ty}} 
 ~ {\ent}{$_\emptyset$}~ \form{\seppred{\code{lseg1}}{x{,}ty}} &  X  & 0.10 & 1 \\
26 & \form{ \seppred{\code{avl}}{x{,}size{,}height{,}bal} }
 ~ {\ent}{$_\emptyset$}~ \form{\seppred{\code{btn}}{x{,}size}} 
  &  X   & 0.08 & 1 \\
27 & \form{ \seppred{\code{tll}}{x{,}ll{,}lr{,}size} }
 ~ {\ent}{$_\emptyset$}~ \form{ \seppred{\code{btn}}{x{,}size}}   
  &  X  & 0.07 & 1 \\
\hline
\end{tabular}
\end{center}\savespace
\caption{Entailment Checking with Auxiliary Lemma Synthesis}
\label{tbl:expr-ent} \saveone \vspace{-2pt} 
\end{table}

We have implemented the proposed ideas 
into an entailment procedure, called {\toolname}\footnote{{\toolname} was initially implemented for the SLCOMP competition \cite{smtcomp:sl:14}.}, starting from SLEEK entailment procedure \cite{Chin:SCP:12}.
{\toolname} invokes Z3 \cite{TACAS08:Moura} to discharge satisfiability over pure formulas.
We have also integrated {\toolname}  into S2 \cite{Loc:CAV:2014} and
extended the new system  to support
a modular verification with partially supplied specification
and incremental specification inference.
In the following, we experiment {\toolname} and the enhanced S2 in entailment and program verification problems.
The experiments were performed on a machine with the Intel i7-960 (3.2GHz)
processor and 16GB of RAM.

\noindent{\bf Entailment Check. }
In Table \ref{tbl:expr-ent},
we evaluate  {\cyclic} and {\toolname} on 
 inductive entailment problems without user-supplied lemmas (i.e. \form{\lemstore{=}\emptyset}).
Ent 1-19 are shape-only problems; they were taken from  
 Smallfoot \cite{Berdine:APLAS05} (Ent 1-5),
and {\cyclic} \cite{Brotherston:CADE:11,Brotherston:APLAS:12} (Ent 6-19).
Ent 20-27 are shape-numerical problems. 
We used \code{sortll} for a sorted list with smallest value \code{min},
and tll for a binary tree whose nodes point to their parents
and leaves are linked as a singly-linked list
 \cite{DBLP:conf/cade/IosifRS13,Loc:CAV:2014}.
 Time is in second.
TO (X) denotes timeout (30s) (not-yet-support, resp.).
The last column \#syn shows the number of lemmas our generated to prove the entailment check.
  The experimental results show that
 {\toolname} can handle a wider range of inductive entailment problems.
 The experiments 
 also demonstrate the efficiency of our implementation;
for 18 problems which both tools successfully verified,
while it took {\cyclic} 8.29 seconds, it took {\toolname} only 3.14 seconds.


\noindent{\bf Modular Verification for Memory Safety. }
We enhance S2 to automatically verify
a wide range of programs with a higher level of correctness and scalability.
In more detail, it automatically verifies those programs in \cite{Nguyen:CAV08}
 (e.g. \code{bubble\_sort}, \code{append} method of list with tail - like
  \code{ll_{last}} predicate in App. \ref{mov.veri.simp})
without any user-supplied lemma.
By generating consequence parallel
separating lemmas,
it also successfully infers shape specifications of
methods which manipulate the last element of a singly-linked
list (i.e. \code{g\_slist\_concat} in \code{gslist.c}) and a
 doubly-linked list (i.e. \code{g\_list\_append} in \code{glist.c}) of GLIB library \cite{glib:13} (See App. \ref{mov.veri.simp}).
By generating equivalence lemmas,
matching a newly-inferred {\ud} predicate
with predefined predicates in S2 is now extended beyond shape-only domain.

We evaluated the enhanced S2 on  
the cross-platform C library
 Glib open source  \cite{glib:13}.
\begin{wrapfigure}{lt}{0.51\textwidth}
 \vspace{-3pt} 
\savespace 
\begin{center}
\begin{tabular}[t]{|c | c |c | c | c | c | c | c |}
\hline
\multirow{2}{*}{} &  \multirow{2}{*}{LOC} &  \multirow{2}{*}{\#Pr} & \multirow{2}{*}{\#Lo} & \multicolumn{2}{|c|}{wo.} & \multicolumn{2}{|c|}{w.}  \\
\cline{5-8}
     &   & & &        \#\okF & sec. & \#\okF & sec. \\
\hline
 gslist.c & 698 & 34 & 18  & 41 &  2.19  & 47 & 2.30\\
 glist.c & 784 &  32 & 19 & 39  &   3.20  & 46 & 3.39 \\
 gtree.c & 1204 &  32 &  8 &  36  &  3.46 & 36 & 3.48\\ 
 gnode.c & 1128 & 38 &  27 & 52   & 7.52 & 53 & 7.58 \\
\hline
\end{tabular}
\end{center}\vspace*{-2mm} \savespace
\caption{Experiments on Glib Programs}
\label{tbl:expr:big}
\savespace 
\end{wrapfigure}
We experimented on
heap-manipulating files, i.e.  
singly-/doubly-linked lists (gslist.c/glist.c),
 balanced binary trees (gtree.c) and N-ary trees (gnode.c).
In Fig.\ref{tbl:expr:big} we list for each file the
 number of lines of code
(excluding comments) LOC, 
number of procedures (
while/for loops) \#Pr (\#Lo). $\#\okF$ and sec. show the number
of procedures/ loops and time (in second)
 for which the enhanced S2 can verify memory safety
without (wo.) and with (w.) the lemma synthesis component.
With the lemma synthesis, the number of procedures/loops
was successfully verified increases from 168 (81\%) to 182 (88\%)
with the overhead of 0.38 seconds.





\section{Related Work and Conclusion} \label{sec.related}
\noindent{\bf Entailment Procedure in SL. }
Past works in SL mainly focus on
developing decision procedures for a
decidable fragment combining linked lists (and trees) with only equality and inequality constraints \cite{Berdine:APLAS05,Cook:CONCUR:2011,pldi:PerezR11,Piskac:CAV:2013,Navarro:APLAS:2013}.
Smallfoot \cite{Berdine:FSTTCS04,Berdine:APLAS05},
 provided
strong semantic foundations and proof system
with frame inference capability for the above fragment.
Some optimization on segment feature for the fragment with linked list
 was presented
in \cite{Cook:CONCUR:2011,pldi:PerezR11,Piskac:CAV:2013,Navarro:APLAS:2013}.
 Recently, Iosif et. al. extended decidable fragment
to restricted {\ud} predicates \cite{DBLP:conf/cade/IosifRS13}.
\cite{Antonopoulos14} presented a comprehensive summary on computational complexity
of deciding entailment in SL with {\ud}
predicates.
Our work, like \cite{Chin:SCP:12,Qiu:PLDI:2013}, targets on an
undecidable SL fragment including
 (arbitrary) {\ud} predicates
and numerical constraints.
Like \cite{Chin:SCP:12,Qiu:PLDI:2013}, we trade completeness for expressiveness.
 Beyond the focus of \cite{Chin:SCP:12,Qiu:PLDI:2013}, we provide inductive reasoning
in SL using lemma synthesis.

\noindent{\bf Lemma Mechanism in SL. }
Lemma is widely used to enhance the reasoning of
heap-manipulating programs. 
For examples, lemmas are used 
as  alternative unfoldings beyond predicates' definitions
 \cite{Nguyen:CAV08,Brotherston:APLAS:12},
 external inference rules \cite{Distefano:2008:OOPSLA},
or intelligent generalization 
to support inductive reasoning
 \cite{Brotherston:CADE:11}.
Unfortunately,
 these systems require user to
supply those additional lemmas that
might be needed for a proof.
In our work, we 
propose to automatically generate lemmas
either dynamically for inductive reasoning
or statically for theorem exploration.

\noindent{\bf Induction Reasoning. }
For a manual and indirect solution for inductive reasoning in SL,
Smallfoot \cite{Berdine:APLAS05}
presented subtraction rules
that are consequent
from a set of lemmas 
of lists and trees.
Brotherston et. al. proposed a {\em top-down} approach
to automate inductive proofs using
 cycle proof \cite{Brotherston:05}. To avoid infinite circular proof search, the
cyclic technique stops expanding
 whenever current sequent is a repetition of a similar proof pattern
detected from historical proof tree.
Cyclic proof was successfully implemented 
 in first-order logic \cite{Brotherston:APLAS:12},
 and separation logic \cite{Brotherston:CADE:11}.
Circularity rule, a similar mechanism to cyclic, was
also introduced in
 matching logic \cite{Rosu:2012:OOPSLA}.
\cite{Chu:PLDI:2015} managed induction by a 
framework with historical proofs.
Our proposal extends these systems with frame inference
and gives better support for 
 modular verification of heap-manipulating programs.


\noindent{\bf Auxiliary Lemma Generation. }
In inductive theorem reasoning,
auxiliary lemmas are generated (and proven) either
{\em top-down} to support inductive proofs (e.g.
IsaPlanner \cite{Ireland:1996}, 
 Zeno \cite{Sonnex:2012:TACAS}
and extension of CVC4 \cite{Andrew:2015:VMCAI})
or {\em bottom-up} to discover theorem 
(e.g.
 IsaCosy \cite{Johansson:2011:AR} and HipSpec \cite{Claessen:2013:CADE}) and \cite{Roy:2007}.
The center of these techniques are
 heuristics to generate useful lemmas for sets of given functions, constants and datatypes.
Typically, while the top-down proposals (i.e. \cite{Sonnex:2012:TACAS})
 suggest new lemmas
by replacing some common sub-term in a stuck goal
by a variable,
the bottom-up proposals
 (i.e. \cite{Claessen:2013:CADE}) generate
lemmas to compute equivalence functions for functional programs.
In our work, we introduce both top-down and bottom-up approaches
into an entailment procedure in SL.  To support inductive entailment
 proofs dynamically,
we generate auxiliary conjectures
with unknown predicates to infer either universal guard
or frame.
To support theorem discovery, we synthesize
equivalence, split/join/reverse and separating conjectures.
This mechanism can be extended to other heuristics to enhance
proofs of a widen class of {\ud} predicates.

\section{Conclusion}
Lemmas have been widely used to enhance the capability of program verification systems.
 However, existing reasoning systems of heap-manipulating programs
 via separation logic
 rely on user to supply additional lemmas that
might be needed for a proof.
In this paper,
we have presented a mechanism for applying, proving and synthesizing lemma
in a SL entailment procedure. We have shown an implementation
that has a higher level of automation and completeness
for benchmarks taken from inductive theorem proving
and software verification sources.
Our evaluation indicates that inductive proofs benefit from both bottom-up
and top-down lemmas generated by our new approach. It also shows
that synthesized lemmas are relevant and helpful to proving
a conjecture. Future work includes 
 extending the incremental inference mechanism to other pure domains, e.g. bag/set domain.
\savespace \savespace \saveone



\bibliography{all}
 \bibliographystyle{plain}

\newpage
 \appendix
\section{Separation Entailment Procedure}\label{entail.rules}


\begin{figure}[hbt]
\begin{center}
\begin{minipage}{0.99\textwidth}
\begin{frameit}
 \vspace{-5pt}
\savespace\[\savespace
\begin{array}{c}\entrulen{INC1}\\
\entailSYN{\sepnodeF{x}{c}{\setvars{v}} {\sep}\D_1}{\lemstore}{\D_2{\wedge}x{=}\nil}{\emptyset}{\heap}
\end{array}
\quad
\begin{array}{c}\entrulen{INC2}\\
\entailSYN{\D_1{\wedge}x{=}\nil}{\lemstore}{\sepnodeF{x}{c}{\setvars{v}} {\sep} \D_2}{\emptyset}{\heap}
\end{array}
\]
\[
\!\frac{\begin{array}{c}\entrulen{M}\\
 \rho{=}[\setvars{v}/\setvars{w}] \quad \pure_{eq} = \text{freeEQ}(\rho)\\
\entailSYN{\D_1{\wedge}\pure_{eq}}{\lemstore}{\D_2\subst{\setvars{v}}{\setvars{w}}}{\RA}{\constr_f}
\end{array}}
{
\entailSYN{\sepnodeF{x}{c}{\setvars{v}} {\sep} \D_1}{\lemstore}{\sepnodeF{x}{c}{\setvars{w}} {\sep} \D_2}{\RA}{\constr_f}
}%
\qquad
\!\frac{\begin{array}{c}\entrulen{PRED-M}\\
 \rho{=}[\setvars{v}/\setvars{w}] \quad \pure_{eq} = \text{freeEQ}(\rho)\\
\entailSYN{\D_1{\wedge}\pure_{eq}}{\lemstore}{\D_2\subst{\setvars{v}}{\setvars{w}}}{\RA}{\constr_f}
\end{array}}
{
\entailSYN{\seppredF{\code{P}}{(r,\setvars{v})} {\sep} \D_1}{\lemstore}{\seppredF{\code{P}}{(r,\setvars{w})} {\sep} \D_2}{\RA}{\constr_f}
}%
\]
\[
\!\frac{\begin{array}{c}\entrulen{LU}\\
 \bigvee \D_{u_i} = \unfold(\seppredF{\code{P}}{\setvars{v}}{\sep}\D_1,0) \\
{\entailSYN{\D_{u_i}{\sep}\D_1}{\lemstore}{\D_2}{\RA_i}{\constr_i}} \quad { i{ =} 1...n}
\end{array}}
{
\entailSYN{\seppredF{\code{P}}{\setvars{v}}{\sep}\D_1}{\lemstore}{\D_2}{\bigwedge \RA_i}{\bigvee \constr_i}
}%
\quad
\!\frac{\begin{array}{c}\entrulen{RU}\\
 \bigvee \D_{f_i} = \unfold(\seppredF{\code{P}}{\setvars{v}}{\sep}\D_2,0) \\
{\entailSYN{\D_1}{\lemstore}{\D_{f_i}{\sep}\D_2}{\RA_i}{\constr_{i}}} \quad { i{ =} 1...n}
\end{array}}
{\entailSYN{\D_1}{\lemstore}{\seppredF{\code{P}}{\setvars{v}}{\sep}\D_2}{\RA_i}{\constr_i}}%
\]
%
%
 \[
\!\frac{\begin{array}{c}\entrulen{ALIAS}\\
{\entailSYN{\sepnodeF{y}{c}{\setvars{v}} {\sep}\heap_1{\wedge}x{=}y{\wedge}\pure_1}{\lemstore}{\D_2}{\RA}{\constr_R}}%
\end{array}}
{\entailSYN{\sepnodeF{x}{c}{\setvars{v}} {\sep}\heap_1{\wedge}x{=}y{\wedge}\pure_1}{\lemstore}{\D_2}{\RA}{\constr_R}}%
%
\quad
\!\frac{\begin{array}{c}\entrulen{XPURE}\\
\pureentail{\xpure(\heap_1){\wedge}\pure_1}
{{\pure_2}}{ \RA}\!\!
\end{array}}
{\entailSYN{\heap_1{\wedge}\pure_1}{\lemstore}{\pure_2}{\RA}{\heap_1{\wedge}\pure_1}}%
\]
\caption{Basic Inference Rules for Entailment Checking}
\label{fig.entail}
 \vspace{-16pt}
\end{frameit}
\end{minipage}
\end{center}
\end{figure}
%
%
%











Entailment procedure between \form{\D_{a}} and \form{\D_{c}} is formalized as follows:
\[
\entailSYN{\D_{a}}{\lemstore}{\D_{c}}{\RA}{\constr_f}
\]

\noindent 
The entailment outputs residual frame  \form{\constr_f} and a set of relational
assumptions \form{\RA}. (For simplicity, we discard footprints and existential quantifiers of consequent in this discussion.)
Inference rules are presented in Fig. \ref{fig.entail}.

To derive a proof for an entailment check,
our system deduces antecedent into two parts (i) relevant one would be subsumed by
models of the consequent; (ii) the rest will be inferred as residual frame.
To do that,
it subtracts (match) heap two sides until
heap in the consequent is empty (via the $\entrulen{*M}$,  $\entrulen{LU}$,  $\entrulen{RU}$ inference rules).
 After that, it semantically check the validity for
the implication of the pure part by using
external SMT solvers and theorem provers (via $\entrulen{XPURE}$ inference rule).
Typically, an entailment check is performed as follows.
\begin{itemize}
\item {\bf Subtracting.} Match up identified 
 heap chains.
       Starting from identified root pointers, the procedure keeps matching all their reachable heaps
with $\entrulen{M}$ and $\entrulen{PRED-M}$ rules.
 The former (latter) rule matches two points-to (user-defined, resp.) predicates in antecedent
and consequent if they have an identified root. 
After that, it unifies corresponding fields of matched roots
 by using auxiliary function $\text{freeEQ}(\rho)$:
\form{\text{freeEQ}([u_i/v_i]^n_{i=1}) = \bigwedge^n_{i=1} \{ u_i = v_i 
\}}.
\item {\bf Unfolding.} Derive alternative 
 heap chains.
 When the procedure is unable to make a progress on matching,
it will look up alternative chains for matching through unfolding heap predicates.
While the unfolding in the antecedent ($\entrulen{LU}$ rule) does cases split,
the unfolding in the consequent ($\entrulen{RU}$ rule) does proof search.
\item {\bf  $\xpure$ Reducing.} Reduce entailment checking on separation logic to implication checking
       on the first order-logic with $\entrulen{XPURE}$ rule.
This reduction was presented in \cite{Loc:2014:S2SAT:TR}.
 When the consequent remains empty heap, e.g. $\emp \wedge \pure_{c}$,
the procedure employs $\entrulen{\xpure}$ inference rule
to decide the entailment result. Firstly, this rules make use of the $\xpure$ reduction to transform the combination of
 remain heaps in the antecedent
and footprints into the first order-logic formula on the combination of pure domains, e.g. $\pure_{a}$.
 Then it checks the implication \form{\pure_{a} \implies \pure_{c}}.
Technically, to perform such implication checking,
 the following satisfiability check is performed:
\form{\satp{\pure_{a} ~{\wedge}~ \neg (\pure_{c})}}. If it returns \code{unsat}, the result of the implication is valid;
 it returns \code{sat}, the result of the implication is invalid; otherwise, the result of the implication is unknown.

\end{itemize}
During the heap chains matching,
aliasing relation on pointers are considered to introduce alternative proofs via
\form{\entrulen{ALIAS}} rule.

\noindent{\bf Instantiation Mechanism. } 
A variable is instantiable if
it is an actual parameter of a {\ud} predicate
instance in the consequent (RHS)
 and is quantifier-free.
 This mechanism is applied for predicate matching rule \cite{Chin:SCP:12} (corresponding rule
 used in \cite{Distefano:2008:OOPSLA} is subtracting) and predicate folding rule \cite{Chin:SCP:12}
(corresponding rule
 used in \cite{Brotherston:CADE:11} is unfolding predicate in RHS).
 Whenever a match of a {\ud} predicate instance occurs,
the entailment procedure binds
its instantiable parameters coming from the RHS
with corresponding variables from the antecedent (LHS)
and moves the equality constraints to the LHS.
Whenever a {\ud} predicate instance in the RHS is unfolded,
our proof system moves pure constraints over instantiable (actual) parameters of
 the unfolded formulas
  to the LHS.
Moreover, this mechanism is proven sound and is able to enhance the completeness of entailment procedure
for a SL fragment including {\ud} predicates with pure properties \cite{Chin:SCP:12}.

\hide{
\subsection{On Complete Entailment Procedure}\label{entail.extra}

\begin{figure*}[hbt]
\begin{center}
\begin{minipage}{0.99\textwidth}
\begin{frameit}
 \vspace{-15pt}
\[
\!\frac{\begin{array}{c}\entrulen{SEG-UNFOLD1}\\
\fresh ~\setvars{y} \quad \code{P} \in \segpred \\
\entailKE{\small \heap}{\small V \cup \{p\} \cup \setvars{y}}{\seppredF{\code{P}}{\self,p,\setvars{v}}{\sep}\seppredF{\code{P}}{p,\nil,\setvars{t}}{\sep}\D_1}{\seppredF{\code{P}}{\self,p,\setvars{y}} {\sep} \seppredF{\code{P}}{p,\nil,\setvars{w}} {\sep}\D_2}{(\constr_{R}, \flow) }
\end{array}}
{\entailVVE{\seppredF{\code{P}}{\self,p,\setvars{v}}\sep\seppredF{\code{P}}{p,\nil,\setvars{t}}{\sep}\D_1}{\seppredF{\code{P}}{\self,\nil,\setvars{w}} {\sep}\D_2}{(\constr_{R}, \flow)}}%
\]
\[
\!\frac{\begin{array}{c}\entrulen{SEG-UNFOLD2}\\
\fresh ~\setvars{y} \quad \code{P} \in \segpred \\
\entailKE{\small \heap}{\small V \cup \{p\} \cup \setvars{y}}{\seppredF{\code{P}}{\self,p,\setvars{v}}{\sep}\seppredF{\code{P}}{p,z,\setvars{t}}{\sep}\sepnodeF{z}{c}{\setvars{v_1}}{\sep}\D_1}{\seppredF{\code{P}}{\self,p,\setvars{y}} {\sep} \seppredF{\code{P}}{p,z,\setvars{w}} {\sep}\D_2}{(\constr_{R}, \flow) }
\end{array}}
{\entailVVE{\seppredF{\code{P}}{\self,p,\setvars{v}}\sep\seppredF{\code{P}}{p,z,\setvars{t}}{\sep}\sepnodeF{z}{c}{\setvars{v_1}}{\sep}\D_1}{\seppredF{\code{P}}{\self,z,\setvars{w}} {\sep}\D_2}{(\constr_{R}, \flow)}}%
\]
\[
 \frac{\begin{array}{c}
\entrulen{RHS-EX1}\\
 \entailKE{\heap}{V{\cup}\{x\}}{\D_1}{([x/v]\D_2)}{(\D_i, \flow)} \quad
 \constr_R{=}\exists~x\,{\cdot}\,\D_i
\end{array}}{\entailVVE{\sepnodeF{x}{c}{\setvars{w}} \sep \D_1}
{(\exists~v\,{\cdot}\,\sepnodeF{v}{c}{\setvars{t}} \sep \D_2)}{(\constr_R, \flow)}}
\]
\caption{Inference Rules on More Complete Entailment Procedure}
\label{fig.entail.extra}
 \vspace{-15pt}
\end{frameit}
\end{minipage}
\end{center}
\end{figure*}

\hide{\begin{verbatim}

rearrange

seg-fold-1
  x::p<\vector{v_1}, y> * y ... z |- x::p<\vector{v_3},z> --> x::p<\vector{v_1}, y> *  y ... z |- x::p<\vector{v_4},y>*y::p<\vector{v_5},z>


seg-fold-2
  x::br-of-p * y ... z |- x::p<\vector{v_3},z> --> x::p<\vector{v_1}, y> *  y ... z |- x::p<\vector{v_4},y>*y::p<\vector{v_5},z>

unfold
  f_i in definition of view
  r_i = f_i * lhs |- rhs
  r = LAND r_i
 ----------------------
  (x::view<...> * lhs |- rhs) --> r.

fold
  f_i in definition of view
  r_i =  lhs |- f_i * rhs
  r = SEARCH r_i
 ----------------------
  ( lhs |- x::view<...> * rhs) --> r.


invalid check

   inv-u(fv(rhs), \top,rhs) ==> true
   lsvl=fv(lhs)  lhs_u = inv-u(lsvl,\top,lhs)
   inv-u(lsvl, lhs_u, rhs) ==> false
 ----------------------
  (lhs |- rhs) --> INVALID.



valid check: \emp heap rhs

   inv-o(lhs) ==> rhs_pure
 ----------------------
  (lhs |- emp /\ rhs_pure) --> VALID.
 

\end{verbatim}}

To enhance the completeness, we introduce two folding rules
 (\form{\entrulen{SEG-UNFOLD1}} and \form{\entrulen{SEG-UNFOLD2}})
on segment predicates and one rule (\form{\entrulen{RHS-EX1}})
 for existential heap matching.

\noindent{\bf Segment Unfold. }
We proposed rules to unfold 
one segment predicate formula in consequent into two segment predicate
formulas at a cutpoint detected in the antecedent.
Compared to \code{Unfolding} inference rules presented in \cite{Berdine:APLAS05,pldi:PerezR11},
our inference rules can be applied for a general class of segment predicates,
instead of only for hard-wired list segment predicate.

For example, \code{skl2} predicate is a segment predicate
with one skip list segment in the inductive rule as follows:
\[
\begin{array}{l}
\code{c_2~\{c_2 ~down; {~c_2 ~ next};\}}\\
\code{\seppred{lseg}{x,ex}} ~\equiv~
\code{\emp} \wedge \code{x{=}ex} 
\quad \vee \code{~\exists ~q \cdot \sepnode{x}{c_2}{\nil{,}q} {\sep}
\seppred{lseg}{d{,}ex}
 \wedge x{\neq}ex};\\
\code{\seppred{skl2}{x,ex}} ~\equiv~
\code{\emp} \wedge \code{x{=}ex} \\
\quad \vee \code{~\exists ~q{,}d \cdot \sepnode{x}{c_2}{d{,}q} {\sep}
\seppred{lseg}{d{,}q} {\sep}
\seppred{skl2}{q{,}ex} \wedge x{\neq}ex};
\end{array}
\]

Consider the following entailment
\[
 \form{\seppredF{\code{skl2}}{x_1,x_2} {\sep} \sepnodeF{x_2}{c_2}{x_3,x_3} * \seppredF{\code{skl2}}{x_3,\nil} \ent \seppredF{\code{skl2}}{x_1,\nil}}
\]
One its successful proof starts by
 the \form{\entrulen{SEG-UNFOLD1}} rule with the \form{x_2} cutpoint of the two segmentation as follows:
\[
\seppredF{\code{skl2}}{x_1,x_2} {\sep} \sepnodeF{x_2}{c_2}{x_3,x_3} * \seppredF{\code{skl2}}{x_3,\nil} \ent \seppredF{\code{skl2}}{x_1,x_2} {\sep}\seppredF{\code{skl2}}{x_2,\nil}
\]

Consequently, two matches of heap chains would be performed:
\begin{enumerate}
\item \form{\seppredF{\code{skl2}}{x_1,x_2}} heap chain of the LHS matches with \seppredF{\code{skl2}}{x_1,x_2} of the RHS (by applying \form{\entrulen{UNFOLD-L}}, \form{\entrulen{UNFOLD-R}}, and \form{\entrulen{XPURE}})
\item \form{\sepnodeF{x_2}{c_2}{x_3,x_3} * \seppredF{\code{skl2}}{x_3,\nil}} heap chain
of the LHS matches with \form{\seppredF{\code{skl2}}{x_2,\nil}} of the RHS 
(by applying \form{\entrulen{UNFOLD-R}}, applying \form{\entrulen{MATCH}} rule
 to match up point-to predicate
\form{\sepnodeF{x_2}{c_2}{x_3,x_3}} , and so on).
\end{enumerate}

\noindent{\bf  Point-to Heap Predicate Matching for Existential Quantifiers. }
For completeness, the formula \form{\exists ~v \cdot \sepnodeF{v}{c}{\setvars{w}}}
in RHS requires a searching over heap domain.
For both completeness and efficiency, we propose a heuristics that
the search space is limited to the defined heaps in the LHS.
This heuristics is implemented in the rule \form{\entrulen{RHS-EX1}}.

For example, consider the entailment:
\form{\sepnodeF{x}{node}{\anon, \nil} {\sep}
\sepnodeF{z}{c_2}{\nil, \nil} {\ent} \exists ~y \cdot
                           \sepnodeF{y}{node}{\anon, \nil}}
which
\form{y} pointer in the RHS is quantified. Our procedure will
match quantified point-to predicate formulas in RHS, like \form{y},
 with a point-to heap predicates formula
 (1) allocated in the LHS and (2) agree on type. In this example,
 \sepnodeF{x}{node}{\anon, \nil} would be matched with
 \form{\exists ~y \cdot \sepnodeF{y}{node}{\anon, \nil}} and
validity of the entailment is proved.



\noindent{\bf Limitation. }
The procedure is based on
 heap chains matching:
 if there are two point-to (or summary heap of one shape predicate)
 predicates at heads of
 LHS and RHS, consumes them;
 if there is one point-to predicate and
one summary predicate at the head,
 {\em unfolds} the summary one and
 hope that it will introduce
a new point-to predicate for {\em matching};
 if there are two summary heap of two different predicates
at heads of LHS and RHS, unfolding the summary LHS and RHS to introduce point-to predicates for matching.
The third case of such {\bf Unfold and Match}
 above suffers from two drawback.
Firstly, this entailment procedure
can not handle {\em loops} on proving:
the same entailment pattern encounters after unfolding LHS, unfolding RHS and matching.
For example, consider the following proof of the entailment 
\form{\seppred{\code{lseg}}{x{,}p} {\sep} \seppred{\code{ll}}{p{,}\nil} ~{\ent}~
  \seppred{\code{ll}}{x{,}\nil}}.

\begin{footnotesize}
\[
\begin{array}{|l|}
\hline
\begin{array}{|lr|}
\hline
... & \\
{\scriptsize 3}\quad \seppred{\code{lseg}}{q_2{,}p} {\sep} \seppred{\code{ll}}{p{,}\nil} {\wedge} q_1{=}q_2~{\ent}~
  \seppred{\code{ll}}{q_2{,}\nil}  & {\bf \scriptstyle MATCH \& ALIAS}\\
{\scriptsize 2}\quad \sepnodeF{x}{c_1}{q_1}{\sep} \seppred{\code{ll}}{q_1{,}\nil} ~{\ent}~\sepnodeF{x}{c_1}{q_2}
  \seppred{\code{ll}}{q_2{,}\nil}  & {\bf \scriptstyle UNFOLD-R}\\
{\scriptsize 1}\quad \sepnodeF{x}{c_1}{q_1}{\sep} \seppred{\code{lseg}}{q_1{,}p}
 {\sep} \seppred{\code{ll}}{p{,}\nil} {\wedge}x{\neq}p ~{\ent}~
  \seppred{\code{ll}}{x{,}\nil} & {\bf \scriptstyle UNFOLD-L}\\
\hline
\end{array}\\
{\scriptsize 0}\quad \seppred{\code{lseg}}{x{,}p} {\sep} \seppred{\code{ll}}{p{,}\nil} ~{\ent}~
  \seppred{\code{ll}}{x{,}\nil} \\

\hline
\end{array}
\]
\end{footnotesize}

with the substitution \form{[x/q_2]},
the entailment at step 3 is identical to the entailment 
at step 0.

Secondly, although
it works well with
shape predicates contain point-to predicate at the head of their branches.
 It would not work with shape predicates contain recursive
summary predicates at the head (to be referred as non-tail recursive shape predicates.).
For instance, with the non-tail recursive
 list segment \code{rlseg}
\[
\begin{array}{l}
\code{\seppred{rlseg}{\self,s}} ~\equiv~
\code{\emp} \wedge \code{\self{=}s} \\
\quad \vee \code{~\exists ~q \cdot
\seppred{rlseg}{\self{,}q} {\sep} \sepnode{q}{c_1}{s} \wedge \self{\neq}p}
;
\end{array}
\]

Let consider the simple
entailment check:
\form{
 \seppred{\code{rlseg}}{x{,}q} {\sep} \sepnode{q}{c_1}{\nil}\ent \seppred{\code{lseg}}{x{,}\nil}
}.
Typically, the procedure unfolds the summary cells in the LHS and RHS. However,
after the unfolding, heads of LHS and RHS are still summary predicates and the procedure
can not match the heads.

To fix those problems, the procedure
should be empowered with  induction proving:
the entailment at step \code{0} would be considered as
 induction hypothesis and the entailment at step \code{3}
would apply that induction and returns a successful checking.
The cyclic technique \cite{Brotherston:CADE:11} is one promising solution.
However that technique did not support frame inference. We will present
a solution with lemma in the next section.

\subsection{Lemma Mechanism for Induction Proving}\label{entail.lemma}
Entailment procedure is desired to provide a mechanism for user to guide the proof search.
In separation logic, technique in \cite{Nguyen:CAV08}
provides a mechanism such that user can interact with
entailment procedure:
User provides lemmas to express
predicates relationships
 and the decision procedure automatically
applies those lemmas
 as alternative predicate unfolding.
In this section, we will present a  mechanism on lemma to support induction proving
on inductive data structures: lemma
is defined as induction hypothesis and
lemma application employs defined lemma as induction proving.

In the following, we summarize the foundation of user-defined lemma and
 lemma application. We show the limitation of 
the existing approach and present our proposal.

\noindent{\bf User-Defined Lemma. }
Lemma can be defined as either weakening or strengthening. For the former,
root pointer have to be specified in a heap predicate formula of the LHS.
For the latter, root pointer have to be specified in a heap predicate formula of the RHS.
\[
\begin{array}{lcl}
\text{Lemma} & \form{\lem} & ::= \text{Left Lemma} \mid  \text{Right Lemma} \\
& \text{Left Lemma} & ::=  ~  \heap_1{\wedge} \pure_1 ~{\rightarrow}~ \exists w^*{\cdot}\heap_2 {\wedge} \pure_2 \\
 & \text{Right Lemma} & ::=  \heap_1 {\wedge} \pure_1~{\leftarrow}~ \exists w^*{\cdot} \heap_2{\wedge}\pure_2 \\
\end{array}
\]

For left lemma, we require that \form{\heap_1{\wedge} \pure_1} must be a heap chain with an explicit root pointer.
Similarly for right lemma, we require that \form{\heap_2{\wedge} \pure_2} must be a heap chain with an explicit root
pointer.

For instance, user can define the lemma \code{lem_1} to express the relationship
of list segment \code{lseg} and full list \code{ll}:
\form{\text{lemma } \text{\bf lem$_1$  } \seppred{\code{lseg}}{\self{,}p{,}n} {\wedge}p{=}\nil \leftrightarrow
              \seppred{\code{ll}}{\self{,}n}},
with the definition of the full list \code{ll}
\[
\begin{array}{l}
\code{\seppred{lls}{\self}{,}n} ~\equiv~
\code{x{=}\nil {\wedge}n{=}0} \\
\quad \vee \code{~\exists ~q,n_1 \cdot \sepnode{\self}{c_1}{q} {\sep}
\seppred{ll}{q{,}n_1} {\wedge}n_1{=}n{+}1}
\end{array}
\]
($\leftrightarrow$ is a shorthand of both left and right lemmas.)

\begin{figure*}[hbt]
\begin{center}
\begin{minipage}{0.78\textwidth}
\begin{frameit}
 \vspace{-12pt}
\[
\!\frac{\begin{array}{c}\entrulen{LAPP-LEFT}\\
 (\heap_l{\wedge} \pure_l ~{\rightarrow}~ \exists w^*{\cdot}\heap_r {\wedge} \pure_r) \in \G \\
\rho ~{=}~ \code{match}(\heap_1{\wedge}\pure, \heap_l{\wedge} \pure_l ) \qquad
 \rho{\neq}\code{err} \qquad \xpure({\heap_1{\sep}\heap_2}){\wedge}\pure \implies (\pure_l {\rho}) \\
{\entailLEM{\G}{\exists w^* {\cdot}(\heap_r{\rho}) {\sep}\heap_2{\wedge}\pure{\wedge} (\pure_r{\rho})}{\heap}{V}{\D_c}{(\constr_R, \flow, \G_o)}}%
\end{array}}
{\entailLEM{\G}{\heap_1 {\sep}\heap_2{\wedge}\pure}{\heap}{V}{\D_c}{(\constr_R, \flow, \G_o)}}%
\]
\[
\!\frac{\begin{array}{c}\entrulen{LAPP-RIGHT}\\
 (\heap_l{\wedge} \pure_l ~{\leftarrow}~ \exists w^*{\cdot}\heap_r {\wedge} \pure_r) \in \G \\
\rho ~{=}~ \code{match}(\heap_1{\wedge}\pure, \heap_r{\wedge} \pure_r ) \qquad
 \rho{\neq}\code{err} \qquad \xpure({\heap_1{\sep}\heap_2}){\wedge}\pure \implies (\pure_r {\rho}) \\
{\entailLEM{\G}{\D_a}{\heap}{V}{\exists w^* {\cdot}(\heap_l{\rho}) {\sep}\heap_2{\wedge}\pure{\wedge} (\pure_l{\rho})}{(\constr_R, \flow, \G_o)}}%
\end{array}}
{\entailLEM{\G}{\D_a}{\heap}{V}{\heap_1 {\sep}\heap_2{\wedge}\pure}{(\constr_R, \flow, \G_o)}}%
\]
\caption{Inference Rules for Lemma Mechanism}
\label{fig.entail.lemma}
 \vspace{-15pt}
\end{frameit}
\end{minipage}
\end{center}
\end{figure*}

\noindent{\bf Automatic Lemma Application. }
To support the new mechanism,
we enhance the entailment formalism as follows:
\[
\form{\entailLEM{\G_i}{\D_A}{\heap}{V}{\D_C}{((\D_R,\flow),\G_0})}
\]
with additional \form{\G_i} (\form{\G_o}) be input (output, respectively) sets of lemmas.

The entailments of basic inference rules only look up and apply lemmas, if applicable, they do not change the set of lemmas.
Those entailments in those rules should be:
 \form{\entailLEM{\textcolor{blue}{\G_i}}{\D_A}{\heap}{V}{\D_C}{((\D_R,\flow),\textcolor{blue}{\G_i}})}.

Inference rules for lemma application is formalized as in Fig. \ref{fig.entail.lemma}:
 $\entrulen{LAPP-LEFT}$ rule for applying left lemma in LHS and
$\entrulen{LAPP-RIGHT}$ rule for applying right lemma in RHS.
The  $\entrulen{LAPP-LEFT}$ rule first do a subtraction of the heap of LHS.
 look up the first heap chain in the LHS of entailment
that forms a valid matching (via \code{match} function) with LHS of a left lemma.
After that, it checks the implication of current context with guard condition \form{\pure_l}.
Finally, it combines the body of lemma into residual formula of the LHS.
The $\entrulen{LAPP-RIGHT}$ rule is similar to, but applied for right lemma on the RHS.

To find a heap chain to match with lemma, the rules
 start from root pointer of the lemma. It look up heap predicates
 of the LHS that match the type of the root, form a initial substitution on root, e.g. \form{\rho_0{=}[(\self/v) ]},
substitute to the lemma and call \code{match} function.
The \code{match}\form{(\D_l,\D_r)} function is implemented by
continuously applying
the matching rules $\entrulen{PTO-MATCH}$ and $\entrulen{SUMMARY-MATCH}$
until heaps of \form{\D_l} and \form{\D_r} are empty.
Subtraction operation in \cite{Nguyen:CAV08} is performed semantically via an entailment procedure with
frame inference capability, it constrains the lemma application on
the procedure capability i.e. if the procedure can not do the subtraction, the lemma application fails.
Since our subtraction operation has been implemented more syntactically
 via rewrite rule (\code{match} function and substitution), our lemma application can overcome the problem above.


\hide{Lemma application procedure \cite{Nguyen:CAV08}
 has been considered as alternative way to unfold predicate:
left lemmas for weakening LHS by unfolding predicates in LHS
(referred as $\entrulen{LAPP-LEFT}$ rule) and right lemmas
for strengthening RHS by
 unfolding predicates in RHS (referred as $\entrulen{LAPP-RIGHT}$ rule).
For example, left lemma application takes a lemma definition,
formula that needs to be applied and a root pointer
 (where the lemma starts to be applied)
as inputs.
The procedure will output a new formula with lemma applied.
First, the procedure automatically cuts
the formula into two parts: one that matches with
 the LHS of the lemma
and another is residual frame.
Second, RHS of the lemma will be combined with the above residual frame
with the RHS of the lemma (with appropriate).
For termination, like unfolding, lemma application is bounded too.}

We illustrate how lemma application is invoked in the following example:
\begin{footnotesize}
\[
\begin{array}{|c|}
\hline
\begin{array}{lr}
 ..... & \\
{\scriptsize 2.2.2}~\seppred{\code{lseg}}{p{,}\nil{,}k}  {\sep} \sepnode{y}{c_1}{\nil} {\wedge} p=p_1 & \\
 \qquad~{\ent}~ \exists~p_1 \cdot~ \seppred{\code{lseg}}{p_1{,}\nil{,}k} & \quad {\bf \scriptstyle MATCH}\\
{\scriptsize 2.2.1}~\sepnode{x}{c_1}{p} {\sep}\seppred{\code{lseg}}{p{,}\nil{,}k}  {\sep} \sepnode{y}{c_1}{\nil} & \\
\qquad ~{\ent}~ \exists~p_1 \cdot~ \sepnode{x}{c_1}{p_1} {\sep}\seppred{\code{lseg}}{p_1{,}\nil{,}k} & \quad {\bf \scriptstyle UNFOLD-R}\\
{\scriptsize 1}~\sepnode{x}{c_1}{p} {\sep}\seppred{\code{lseg}}{p{,}\nil{,}k}  {\sep} \sepnode{y}{c_1}{\nil} ~{\ent}~
 \seppred{\code{lseg}}{x{,}\nil{,}k{+}1} & \quad {\bf \scriptstyle LAPP-RIGHT}
\end{array}\\
\hline
{\scriptsize 0}~\sepnode{x}{c_1}{p} {\sep}\seppred{\code{lseg}}{p{,}\nil{,}k}  {\sep} \sepnode{y}{c_1}{\nil} ~{\ent}~
 \seppred{\code{ll}}{x{,}k{+}1} \\
\qquad \yields ( \sepnode{y}{c_1}{\nil} {\wedge} p=p_1 , \text{valid})\\
\hline
\end{array}
\]
\end{footnotesize}

The lemma mechanism enhances the completeness of entailment procedures.
However, it has two major disadvantages:
\begin{enumerate}
\item Automation: lemmas have to be provided manually.
\item Efficiency: lemma application always is considered for all proof searches,
 even in inapplicable scenarios and thus significantly increases search space.
\end{enumerate}

A better design of lemma mechanism should automatically
 synthesize and apply lemmas on-demand.

\subsection{On Induction Proving}
\label{entail.cycle}
We have discussed entailment procedure on heap predicate formulas from
the same predicate definition.
To support entailment check among different predicate definitions,
 we will extend the entailment procedure to inductive reasoning in this section.
We will refer those inference rules in Fig. \ref{fig.entail} and Fig. \ref{fig.entail.extra} above as basic rules.

One promising solution to check entailment among different predicates
is lemma mechanism \cite{Nguyen:CAV08}.
This mechanism leads to a foundation of
induction proving for data structures using separation logic.

\noindent{\bf Induction Proving. }
Entailment involving inductive predicates may contains cycle proof and
thus lead to infinite runs. Cyclic proof \cite{Brotherston:APLAS:12}
is a promising technique to handle such entailment.
This technique detects the start point of a cycle proof, called a cycle {\em checkpoint}.
The entailment at that point is considered as induction hypothesis. When
the proof makes progress (e.g. on heap size) and can link back to the checkpoint (with generalization),
the induction hypothesis is applied; the entailment is proved as success.
However, cyclic technique suffers from at least
two problems. First there is no chance for its \code{Cut} rule to
automatically employ user's guide.
As discussed by the authors of \cite{Brotherston:APLAS:12},  for the automation of
 the \code{Cut} rule, cyclic technique may
need to support
 lemma application to make use of user's guide via user-defined
 lemmas.
Second, when linking back,
induction hypothesis always applies for whole present entailment, but not apart of
LHS of the entailment. As such, it limits the chance to apply induction hypothesis.

\noindent{\bf Lemma Synthesis. }

\begin{figure*}[hbt]
\begin{center}
\begin{minipage}{0.98\textwidth}
\begin{frameit}
 \vspace{-2pt}
\[
\renewcommand{\arraystretch}{1.2}
\begin{array}[t]{c}
\entrulen{LSYN-2PRED}\\
\D_1'{=} \chain{\seppred{\code{P}}{r{,}\bar{w_1}} {\sep} \heap_1 {\wedge} \pure_1}{r}{\setvars{w}_2} \\      
 \lem {=} \synlemma{\G_i}{\seppred{\code{P}}{r{,}\bar{w_1}}}{\D_1'}
  {\seppred{\code{Q}}{r{,}\bar{w_2}} {\wedge}\project{\pure_2}{r{::}\bar{w_2}}}{\llem}
 \qquad \lem ~{\not=}~ \code{\lemerr}\\
\D_2' {=} \lemapp({\seppred{\code{Q}}{r{,}\bar{w_2}} {\sep}\heap_2 {\wedge} \pure_2}, \lem) 
\qquad \entailLEM{(\G_i \cup \{\lem \})}
{\seppred{\code{P}}{r{,}\bar{w_1}} {\sep} \heap_1 {\wedge} \pure_1}{\heap}{V}{\D_2'}{((\constr_R,\flow),\G_o)}\\
\hline
\entailLEM{\G_i}{\seppred{\code{P}}{r{,}\bar{w_1}} {\sep} \heap_1 {\wedge} \pure_1}{\heap}{V}
{\seppred{\code{Q}}{r{,}\bar{w_2}} {\sep} \heap_2 {\wedge} \pure_2}{((\constr_r,\flow),\G_o)}
\end{array}
\]
\hide{\[
\renewcommand{\arraystretch}{1.2}
\begin{array}[t]{c}
\entrulen{LSYN-2PRED}\\
 \lem {=} \synlemma{\G_i}{\seppred{\code{P}}{r{,}\bar{w_1}}}{\project{\pure_1}{r{::}\bar{w_1}}}
  {\seppred{\code{Q}}{r{,}\bar{w_2}}} {\project{\pure_2}{r{::}\bar{w_2}}}{\rlem}
 \qquad \lem ~{\not=}~ \code{\lemerr}\\
\D_2' {=} \lemapp({\seppred{\code{Q}}{r{,}\bar{w_2}} {\sep}\heap_2 {\wedge} \pure_2}, \lem) 
\qquad \entailLEM{(\G_i \cup \{\lem \})}
{\seppred{\code{P}}{r{,}\bar{w_1}} {\sep} \heap_1 {\wedge} \pure_1}{\heap}{V}{\D_2'}{((\constr_R,\flow),\G_o)}\\
\hline
\entailLEM{\G_i}{\seppred{\code{P}}{r{,}\bar{w_1}} {\sep} \heap_1 {\wedge} \pure_1}{\heap}{V}
{\seppred{\code{Q}}{r{,}\bar{w_2}} {\sep} \heap_2 {\wedge} \pure_2}{((\constr_r,\flow),\G_o)}
\end{array}
\]}
\[
\renewcommand{\arraystretch}{1.2}
\begin{array}[t]{c}
\entrulen{LSYN-FOLD}\\
\setvars{w_2} \cap \setvars{w_3} \neq \emptyset \qquad
\pure_1'{=}\project{\pure_1}{\{r,p,\setvars{w}_1,\setvars{w}_2\}} \qquad
\pure_2'{=}\project{\pure_2}{r{::}\bar{w_3}}
\\
   \lem {=} \synlemma{\G_i}{\seppred{\code{Q}}{r{,}\bar{w_3}}}
{\sepnodeF{r}{c}{p{,}\setvars{w_1}} {\sep} \seppred{\code{P}}{p{,}\bar{w_2}} {\wedge}\pure_1}
   {\seppred{\code{Q}}{r{,}\bar{w_3}}{\wedge}\pure_2'}{\rlem} \qquad \lem ~{\not=}~ \lemerr \\
 \D_2' {=} {\lemapp(\seppred{\code{Q}}{r{,}\bar{w_3}} {\sep} \heap_2 {\wedge} \pure_2 {,}\lem)} \\
\entailLEM{(\G_i \cup \{\lem \})}{\sepnodeF{r}{c}{p{,}\setvars{w_1}} {\sep} \seppred{\code{P}}{p{,}\bar{w_2}}{\sep} \heap_1 {\wedge} \pure_1}{\heap}{V} {\D_2'}
 {((\constr_R,\flow),\G_o)}\\
\hline
\entailLEM{\G_i}{\sepnodeF{r}{c}{p{,}\setvars{w_1}} {\sep} \seppred{\code{P}}{p{,}\bar{w_2}} {\sep}
                 \heap_1 {\wedge} \pure_1}{\heap}{V}{
\seppred{\code{Q}}{r{,}\bar{w_3}} {\sep} \heap_2 {\wedge} \pure_2}
{((\constr_R,\flow),\G_o)}
\end{array}
\]
\[
\renewcommand{\arraystretch}{1.2}
\begin{array}[t]{c}
\entrulen{LSYN-UNFOLD}\\
   \lem {=} \synlemma{\G_i}{ \seppred{\code{P}}{r{,}\bar{w_1}}{\wedge}\project{\pure_1}{r{::}\bar{w_1}}}
{\seppred{\sepnodeF{r}{c}{p{,}\setvars{w_3}} {\sep} \code{Q}}{p{,}\bar{w_2}}} {\pure_2}{\llem} \qquad \lem ~{\not=}~ \lemerr \\
  {\D_1'} {=} \lemapp( \seppred{\code{P}}{r{,}\bar{w_1}} {\sep}  \heap_1 {\wedge} \pure_1, \lem ) \\
\entailLEM{(\G_i \cup \{\lem \})}{ \D_1'}
          {\heap}{V}
{\seppred{\sepnodeF{r}{c}{p{,}\setvars{w_3}} {\sep} \code{Q}}{p{,}\bar{w_2}} {\sep}\heap_2 {\wedge} \pure_2}{((\constr_R,\flow),\G_o)}\\
\hline
\entailLEM{\G_i}{ \seppred{\code{P}}{r{,}\bar{w_1}} {\sep} \heap_1 {\wedge} \pure_1}
{\heap}{V}
{\seppred{\sepnodeF{r}{c}{p{,}\setvars{w_3}} {\sep} \code{Q}}{p{,}\bar{w_2}} {\sep} \heap_2 {\wedge} \pure_2}{((\constr_R,\flow),\G_o)}
\end{array}
\]
\caption{Inference Rules for Induction Proving}\label{fig.entail.cycle}
 \vspace{-15pt}
\end{frameit}
\end{minipage}
\end{center}
\end{figure*}

\hide{
\begin{figure}[hbt]
\begin{center}
\begin{minipage}{0.98\textwidth}
\begin{frameit}
 \vspace{-2pt}
\[
\renewcommand{\arraystretch}{1.2}
\begin{array}[t]{c}
\entrulen{LSYN-2PRED}\\
\D_1'{=} \chain{\seppred{\code{P}}{r{,}\bar{w_1}} {\sep} \heap_1 {\wedge} \pure_1}{r}{\setvars{w}_2} \\
 \lem {=} \synlemma{\G_i}{\seppred{\code{P}}{r{,}\bar{w_1}}}{\D_1'}
  {\seppred{\code{Q}}{r{,}\bar{w_2}}} {\project{\pure_2}{r{::}\bar{w_2}}}{\rlem}
 \qquad \lem ~{\not=}~ \code{\lemerr}\\
\D_2' {=} \lemapp({\seppred{\code{Q}}{r{,}\bar{w_2}} {\sep}\heap_2 {\wedge} \pure_2}, \lem) 
\qquad \entailLEM{(\G_i \cup \{\lem \})}
{\seppred{\code{P}}{r{,}\bar{w_1}} {\sep} \heap_1 {\wedge} \pure_1}{\heap}{V}{\D_2'}{((\constr_R,\flow),\G_o)}\\
\hline
\entailLEM{\G_i}{\seppred{\code{P}}{r{,}\bar{w_1}} {\sep} \heap_1 {\wedge} \pure_1}{\heap}{V}
{\seppred{\code{Q}}{r{,}\bar{w_2}} {\sep} \heap_2 {\wedge} \pure_2}{((\constr_r,\flow),\G_o)}
\end{array}
\]
\caption{Inference Rules for Induction Proving}\label{fig.entail.cycle}
 \vspace{-15pt}
\end{frameit}
\end{minipage}
\end{center}
\end{figure}
}

We are fixing the problems of existing lemma mechanism in \cite{Nguyen:CAV08} and cyclic proof \cite{Brotherston:APLAS:12}.
We propose a novel lemma synthesis mechanism such that it automatically
(1) detects and synthesizes induction hypothesis as lemmas
and (2) links back current entailment to induction hypothesis
 to form a cycle proof by lemma application mechanism.
As such, lemmas are synthesized automatically and applied on-demand.
Moreover, with automatic lemma application mechanism,
 we can support lemma application to employ user's guide.
Furthermore, with frame inference capability, induction hypothesis is
apart of entailment and it can also apply to apart of the cycle entailment.

To support the new mechanism,
we enhance the entailment formalism as follows:
\[
\form{\entailLEM{\G_i}{\D_A}{\heap}{V}{\D_C}{((\D_R,\flow),\G_0})}
\]
with additional \form{\G_i} (\form{\G_o}) be input (output, respectively) sets of lemmas.

The entailments of basic inference rules only look up and apply lemmas, if applicable, they do not change the set of lemmas.
Those entailments in those rules should be:
 \form{\entailLEM{\textcolor{blue}{\G_i}}{\D_A}{\heap}{V}{\D_C}{((\D_R,\flow),\textcolor{blue}{\G_i}})}.


Inference rules supporting lemma synthesis at checkpoints is presented in Fig. \ref{fig.entail.cycle}.
The $\entrulen{LSYN-2PRED}$ rule handles the scenario when the LHS and RHS are at summary predicates.
Before unfolding predicates (i.e. make a progress),
the procedure invokes \code{syn} function for detecting possible cycle entailment:
\begin{itemize}
\item \code{syn} captures a checkpoint by synthesizing a lemma as induction hypothesis;
\item \code{syn} makes progresses until the entailment is proved
successfully and the induction hypothesis (the synthesized lemma) is applied at least once.
\end{itemize}
If the cycle entailment is proved successfully, the lemma is synthesized
and applied to the original entailment to cut the cycle entailment up
 (by \code{lem\_app function}).
We note that in this rule, we add a heuristic that the lemma is synthesized only when the two predicates
are not identical.
\hide{ Similarly, checkpoints are also captured before unfolding (folding)
via $\entrulen{LSYN-UNFOLD}$ ($\entrulen{LSYN-FOLD}$, resp.) rule.}

\begin{footnotesize}
\[
\begin{array}{|c|}
\hline
\begin{array}{lr}
 ..... & \\

\end{array}\\
\hline
{\scriptsize 0}~ \seppred{\code{lsegs}}{x{,}p{,}k}  {\sep} \seppred{\code{lls}}{p{,}\nil{,}l} ~{\ent}~
 \seppred{\code{lls}}{x{,}k{+}l} \\
\qquad \yields ( \sepnode{y}{c_1}{\nil} {\wedge} p=p_1 , \text{valid})\\
\hline
\end{array}
\]
\end{footnotesize}

}

\section{Theorem Exploration} \label{app.pred.trans}
\subsection{Generating Reverse/Split/Join Lemmas} \label{app.pred-split}
\noindent{\bf Step 1.} This subsection explores
theorem over segment predicates as follows.
\begin{defn}[Segment Predicate]
A predicate \form{\seppred{\code{SP}}{r,\setvars{v},s}}
 is a segment predicate if r is a root parameter
and for any base formula \form{\D} which is derived by unfolding  
 \form{\seppred{\code{SP}}{r,\setvars{v},s}}, \form{s} is a leaf pointer
reached from \form{r}. We will refer \form{s} as a segment parameter.
\end{defn}
For instance, linked-list segment predicate
with size property
is defined as follows:
\savespace
\[\savespace
\begin{array}{l}
\form{\defpred~\seppred{\code{glsegn}}{\self{,}s{,}n}} {\equiv}
\form{\emp} {\wedge} \form{\code{\self}{=}s {\wedge} n{=}0}
~ {\vee} \form{~\exists ~q {\cdot}~ \sepnode{\code{\self}}{c_1}{q} {\sep}
\seppred{\code{glsegn}}{q{,}s{,}n{-}1} }
;
\end{array}
\]
 The predicate \code{glsegn} above
may be an acyclic list, or a complete cyclic list,
or a lasso (an acyclic fragment followed by a cycle).
The {\em acyclic} list segment predicate \code{lsegn} (Sec. \ref{sec:motivate})
 is a special segment predicate.
Tree segment predicates can be found in \cite{Brotherston:CADE:11}.

A {\ud} predicate is syntactically classified as segment predicate
if it has  one root parameter \form{r} and one segment parameter \form{s}
such that \form{s} is a leaf pointer which is reached from \form{r}.
A {\ud} predicate is syntactically classified as {\em acyclic} segment predicate
if it is 
 a segment predicate
and the formula \form{r{\neq}s} occurs in all inductive branches.

\noindent{\bf Step 2. }
Reverse lemmas explore relations between
reverse directions of linked heaps i.e.
forwardly and backwardly linked list from \form{\self} parameter
to segment parameter in inductive branches of segment predicates.
Our implementation for reverse lemmas currently restricts for
reachable heaps linked by points-to predicates and segment predicate instances.
With a segment predicate \form{\seppred{\code{Q}}{\self,\setvars{w},s}},
for each inductive branch \form{\exists \setvars{w}_i{\cdot}\D_i},
 reverse linked heaps
\form{\exists \setvars{w}_i{\cdot}\D_{i_r}} of  
is examined as follows:
(i)  mark reachable heaps, a set of points-to predicates and segment predicate instances
 from \form{\self} to \form{s};
(ii)  swap \form{\self} and \form{s}. Now, \form{s}  is a root variable of
either a points-to predicate
or  a segment predicate instance;
(iii) starting from the heap predicate with \form{s},  reverse
the reachable heaps following the links. (For each points-to predicate,
swap points-to variable with downstream field variable of the links.
For each segment predicate instance, swap root parameter and segment parameter);
(iv) keep the rest of \form{\D_i} unchanged.

\noindent Reverse conjectures are initially generated over reachable
heaps as: \form{\exists \setvars{w}_i{\cdot} \D_{i_r} {\rightarrow} \seppred{\code{Q}}{\self{,}\setvars{w}{,}s}}.

For example, with segment list \code{glsegn} above, we generate
the following lemma:
\form{\code{lemma~rev_1} {~\exists ~q {\cdot}~ \sepnode{q}{c_1}{s} {\sep}
\seppred{\code{glsegn}}{\code{\self}{,}q{,}n{-}1} ~{\rightarrow}~ \seppred{\code{glsegn}}{\code{\self}{,}n{,}s}}}.

 Heap-only conjecture to explore join relation
for the segment predicate \form{\seppred{\code{P}}{x,\setvars{w},s}}
(\form{x} is a root parameter and \form{s} is a segment parameter)
is  generated as:
\[ \form{\exists z{,}\setvars{w}_1,\setvars{w}_2 \cdot \seppred{\code{P}}{x,\setvars{w}_1,z}{\sep}
\seppred{\code{P}}{z,\setvars{w}_2,s} \rightarrow {\exists}\setvars{w}{\cdot}
\seppred{\code{P}}{x,\setvars{w},s}}\]

Two heap-only conjectures to explore join relation
for the acyclic segment predicate \form{\seppred{\code{Q}}{x,\setvars{w},s}}
of data type \code{data ~ c \{f_i{:}v_i;\}}
are  generated as:
\[
\form{\exists z,\setvars{w}_1,\setvars{w}_2 \cdot \seppred{\code{Q}}{x,\setvars{w}_1,z}{\sep}
\seppred{\code{Q}}{z,\setvars{w}_2,\nil} \rightarrow {\exists}\setvars{w}{\cdot}
\seppred{\code{Q}}{x,\setvars{w},\nil}}
\]
 and
\[
\form{\exists z,\setvars{w}_1,\setvars{w}_2 \cdot \seppred{\code{Q}}{x,\setvars{w}_1,z}{\sep}
\seppred{\code{Q}}{z,\setvars{w}_2,s}{\sep}\sepnodeF{s}{c}{\setvars{w}_3} \rightarrow {\exists}\setvars{w},\setvars{w}_3{\cdot}
\seppred{\code{Q}}{x,\setvars{w},s}\sepnodeF{s}{c}{\setvars{w}_3}}
\]

Similarly, {\em split} heap-only conjecture
 \form{\D \rightarrow \D_1{\sep}\D_2}
 is generated as a opposite form of the corresponding join heap-only conjecture \form{\D_1{\sep}\D_2 \rightarrow \D}.

\subsection{Generating Separating Lemmas}\label{app.pred.separation}

\noindent{\bf Step 1. }
This subsection explores relations over {\ud} predicates
including either parallel
or consequence separating parameters.
Two parameters of a predicate are {\em parallel} separating
if they are both root parameters (e.g. those of the predicate \code{zip}, Sec. \ref{sec.lemprove}).
Two parameters of a predicate are {\em consequence} separating
if one is root parameter and another parameter
is internal variable reachable from the root
in all base formulas derived by unfolding the predicate
(e.g. those of the predicate \code{U_{post}}, Sec. \ref{mov.veri}).
We generate these separating lemmas to explicate separation globally.
As a result, the separation of actual parameters
 is visible from analyses.
This visible separation enables strong updates in modular heap analysis
or
frame inference in modular verification.

\noindent{\bf Step 2. } Suppose
\form{r_1}, \form{r_2} are consequence or parallel parameters in
\form{\seppred{\code{Q}}{r_1,r_2,\setvars{w}}},
heap conjecture is generated as:
\form{\seppred{\code{Q}}{r_1,r_2,\setvars{w}} \rightarrow
\seppred{\code{Q_1}}{r_1} {\sep} \seppred{\code{Q_2}}{r_2}{\sep}\seppred{\code{Q_3}}{\setvars{w}}}.

For example, the \code{zip} predicate is suggested to split through the following  parallel separating conjecture:
\form{\code{lemma~para~}\seppred{\code{zip}}{\self{,}r_2}
~{\rightarrow} ~
 \seppred{\code{Q_1}}{\self} {\sep}  \seppred{\code{Q_2}}{r_2}}.



\section{More Examples}
\subsection{Universal Lemma Synthesis}\label{mov-pure}

To illustrate the lemma synthesis, consider the following entailment check \code{E_1}
\savespace \[\savespace 
\exists k{\cdot}\seppred{\code{lln}}{x,n} 
{\wedge}
n{\geq}k {\wedge}k{\geq}0{\wedge}i{=}k{\wedge}j{=}n{-}k~{\ent}_{\emptyset}~ {\exists} p {\cdot} \seppred{\code{lsegn}}{x{,}p{,}i} {\sep}
\seppred{\code{lln}}{p,j}
\]
We define the \code{lln} (\code{lsegn}) predicate to describe an
 acyclic singly-linked list
\form{\nil}-terminated (segment, respectively) over the data type \code{c_1}
 with size property \code{n} as follows:\savespace
\begin{small}\[\savespace
\begin{array}{lr}
\begin{array}{l}
\form{\defpred~ \seppred{\code{lln}}{\self{,}n}} {\equiv}
\form{\emp} {\wedge} \form{\code{\self}{=}\nil {\wedge} n{=}0} \\
~ {\vee} \form{~\exists ~q {\cdot}~ \sepnode{\code{\self}}{c_1}{q} {\sep}
\seppred{\code{lln}}{q{,}n{-}1} };  \\
\end{array}~&~
\begin{array}{l}
\form{\defpred~\seppred{\code{lsegn}}{\self{,}s{,}n}} {\equiv}
\form{\emp} {\wedge} \form{\code{\self}{=}s {\wedge} n{=}0} \\
~ {\vee} \form{~\exists ~q {\cdot}~ \sepnode{\code{\self}}{c_1}{q} {\sep}
\seppred{\code{lsegn}}{q{,}s{,}n{-}1} {\wedge} \form{\self{\neq}s}}; \\
\end{array}
\end{array}
\]
\end{small}
 whereas
  \form{\code{struct~ c_1\{{c_1{\sep} ~ next};\}}}.
\code{E_1} verifies that the list \form{x} can be split into two smaller list segments.
(The list y is the residue.)
To derive a proof for
\code{E_1}, our system automatically generates 
the following auxiliary conjecture with universal guards
and unknown predicate \form{P}:
\savespace
\[\savespace
\begin{array}{l}
\form{\code{lemma} ~\code{sp}~ {\forall} n{,}a{,}b {\cdot}
  \seppred{\code{lln}}{\self{,}n} {\wedge}  P(n{,}a{,}b)
\rightarrow
\exists p \cdot \seppred{\code{lsegn}}{\self{,}p{,}a} {\sep}
 \seppred{\code{lln}}{p{,}b}}.\\
\end{array}\]
and its reverse conjecture
\savespace \[\savespace
\begin{array}{l}
\form{\code{lemma} ~\code{jn}~ \forall n{,}a{,}b {\cdot} 
\exists p {\cdot} \seppred{\code{lsegn}}{\self{,}p{,}a} {\sep}
 \seppred{\code{lln}}{p,b}
\rightarrow
\seppred{\code{lln}}{\self,n} {\wedge} P(n{,}a{,}b)}.
\end{array}\]
Our system will prove the validity of
one lemma and infer definition of \form{P}, simultaneously.
After that, this inferred definition of \form{P}
 is substituted into another lemma before  this lemma is proven.
As \code{jn} has more predicates,
it may need more case splits.
Thus we choose \code{jn} for inference
since it would generate
more relational constraints and our system can obtain more precise
definition of \code{P}.
To prove \code{jn},
our lemma proving component unfolds the predicate  \form{\seppred{\code{lsegn}}{\self{,}p{,}a}}
 in the antecedent and
generates the two subgoals \code{E_2} and \code{E_3} as 
\begin{small}
\savespace\[\savespace
\begin{array}{ll}
\code{E_2{:}} & {\forall} n{,}a{,}b {\cdot}
  \exists p {\cdot} \seppred{\code{lln}}{p,b}{\wedge}\self{=}p{\wedge}a{=}0
~{\ent_{\{\code{jn}\}}}~
\seppred{\code{lln}}{\self,n}{\wedge} P(n{,}a{,}b)
\\
\code{E_3{:}} & {\forall} n{,}a{,}b 
 \exists p{,}q_1{,}a_1 {\cdot} \sepnode{\self}{c_1}{q_1} {\sep} \seppred{\code{lsegn}}{q_1{,}p{,}a_1} {\sep}
 \seppred{\code{lln}}{p,b} {\wedge}a{=}a_1{+}1~ {\ent_{\{\code{jn}\}}}
 \seppred{\code{lln}}{\self,n}{\wedge}P(n{,}a{,}b)\\
\end{array}
\]\end{small}
Inspired by cyclic proof systems \cite{Brotherston:05,Brotherston:CADE:11,Rosu:2012:OOPSLA,Chu:PLDI:2015},
our system has employed the lemma \code{jn}
as induction hypothesis for proving \code{E_2} and \code{E_3}.
For \code{E_2}, we subtract (match) the predicate \code{lln} pointed by
\form{\self} in both sides, instatiate \form{n} and generate the following assumption
to successfully prove the rest of RHS:
\savespace\[\savespace
\begin{array}{l}
\RA_2{:~}a{=}0{\wedge}n{=}b{\wedge}b{\geq}0 ~{\imply}~ P(n{,}a{,}b)\\
\end{array}
\]
For \code{E_3}, we unfold the predicate \code{lln} pointed by
\form{\self} in RHS (recursive case), subtract points-to predicate pointed by
\form{\self} in both sides, apply lemma \code{jn}
 and generate the following assumption
to successfully prove the rest of RHS:
\savespace\[\savespace
\begin{array}{l}
\RA_3{:~}
n{=}n_1{+}1{\wedge}a{=}a_1{+}1{\wedge}P(n_1{,}a_1{,}b){\wedge}n_1{\geq}0{\wedge}a_1{\geq}0{\wedge}b{\geq}0 ~{\imply}~ P(n{,}a{,}b)
\end{array}
\]
Using a fixed point computation (i.e. FixCalc \cite{Popeea:ASIAN06}) to solve \form{\RA_2{\wedge}\RA_3},
a definition of \form{P} can be derived as \form{P(n{,}a{,}b) {\equiv} n{=}a{+}b{\wedge}n{\geq}b{\wedge}b{\geq}0}.
The lemma \code{jn} is synthesized as:
\savespace\[\savespace
\code{lemma} ~\code{jn}~ {\forall} n{,}a{,}b {\cdot}
 {\exists} p {\cdot} \seppred{\code{lsegn}}{\self{,}p{,}a} {\sep}
 \seppred{\code{lln}}{p,b}
\rightarrow
 \seppred{\code{lln}}{\self,n} {\wedge} n{=}a{+}b
{\wedge}n{\geq}b{\wedge}b{\geq}0
\]
Moreover, our system also successfully verifies the lemma
\code{sp} (substituted with \form{P}) as:
\savespace \saveone\[\savespace
\code{lemma} ~\code{sp}~ {\forall} n{,}a{,}b {\cdot}
  \seppred{\code{lln}}{\self,n} {\wedge} n{=}a{+}b
{\wedge}n{\geq}b{\wedge}b{\geq}0
\rightarrow
\exists p \cdot \seppred{\code{lsegn}}{\self{,}p{,}a} {\sep}
 \seppred{\code{lln}}{p,b}
\]
Now, \code{jn} and \code{sp} can be soundly applied
for upcoming proof search.  By applying
 the lemma 
 \code{sp}, 
our entailment procedure
can prove the validity of \code{E_1}
and inferring the residue as: \form{\D_{\code{frame}}{\equiv}
\emp{\wedge}
n{\geq}k {\wedge}k{\geq}0{\wedge}i{=}k{\wedge}j{=}n{-}k{\wedge}i{=}a'{\wedge}j{=}b'}.

 \subsection{Modular Verification with Last Element}\label{mov.veri.simp}
 \begin{wrapfigure}{l}{0.42\textwidth}\savespace
\begin{center}
\begin{small}
  \savespace
\savespace \[
\begin{array}{ll}
1 & \code{c_1{\sep}~append\_shape(c_1~x,  ~c_1~y)} \\
2 & /{*} \form{{\bf\requires}~ \seppred{\code{ll}}{x}{\sep}
 \seppred{\code{ll}}{y} {\wedge}x{\neq}\nil}\\
3 & \form{~~~~ {\bf\ensures} ~ \seppred{\code{ll}}{\res}} */ \\
4 & \code{\{  c_1{\sep}t{=}x;}  \\
5 & \code{~~ while(t{\der}next)~ t{=}t{\der}next;} \\
6 & \code{~~ t{\der}next{=}y;} \\
7 & \code{~~ return~x};\}
\end{array}
\] \savespace 
\savespace 
 \savespace \savespace
\caption{Code of method \code{append\_shape}.}
\label{fig.mov.append.simp}
\end{small}
\end{center}
\end{wrapfigure}

To illustrate how our proof system can support
induction reasoning together with
complex frame inference,
consider the {\em modular} verification of the method \code{append\_shape}
in Fig. \ref{fig.mov.append.simp}.
This method appends a list pointed by \code{y}
to the end of a list pointed by \code{x}.
 The user provides its pre-post specification (lines 2-3)
and the predicates \code{ll}
\savespace
\[\savespace
\begin{array}{l}
\form{\defpred~ \seppred{\code{ll}}{\self}} {\equiv}
\form{\emp} {\wedge} \form{\code{\self}{=}\nil } \\
~ {\vee} \form{~\exists ~q {\cdot}~ \sepnode{\code{\self}}{c_1}{q} {\sep}
\seppred{\code{ll}}{q} };  \\
 \end{array}\]
 whereas
  \form{\code{struct~ c_1\{{c_1{\sep} ~ next};\}}}.
The \code{while} loop is annotated with the natural invariant as follows:
\form{{\bf\requires} ~ 
 \seppred{\code{ll}}{t}{\wedge}t{\neq}\nil ~ {\bf\ensures} ~ \seppred{\code{ll\_last}}{t{,}t'} },
whereas \code{t'} is the value of \code{t} after the loop
and the predicate \code{ll\_last} is supplied as
  \footnote{Indeed, this invariant is the outcome
 of the state-of-the-art shape analysis tools like \cite{Brotherston-Gorogiannis:SAS:14,Loc:CAV:2014} when
they are used for invariant inference.}
\savespace\[\savespace
\begin{array}{l}
\form{\defpred~\seppred{\code{ll\_last}}{\self{,}s}} {\equiv}
 \form{\sepnode{\code{\self}}{c_1}{\nil}{\wedge}\code{\self}{=}s} \\
~ {\vee} \form{~\exists ~q {\cdot}~ \sepnode{\code{\self}}{c_1}{q} {\sep}
\seppred{\code{ll\_last}}{q{,}s}}; \\
\end{array}
\]

As a (bottom-up) modular verification,
the loops are verified prior to
the verification of the method \code{check}; and the correctness of a method
is reduced to the validity of appropriate verification conditions generated.
Our system generates verification conditions to ensure absence of memory
errors (no null dereference, no double free and no memory leak),
 validity of  functional calls/loops via  compositional pre-/post- conditions
 and post-conditions holding. 

The most challenging step to verify this example
is the proving of absence of null dereference
at line 6.
The symbolic state is computed before line 6 is
\savespace\[\savespace
\form{\seppred{\code{ll\_last}}{t{,}t'}{\sep} \seppred{\code{ll}}{y} {\wedge}t{=}x{\wedge}x{\neq}\nil}
\]
 For memory safety at line 6, our system generates
the following proof obligation
\savespace\[\savespace
\form{\form{\seppred{\code{ll\_last}}{t{,}t'}{\sep} \seppred{\code{ll}}{y} {\wedge}t{\neq}\nil}}
~ {\ent_{\emptyset}}~
 \sepnode{\code{t'}}{c_1}{q}
\]
Since the information of \code{t'} is deeply
embedded in the base case of predicate \code{ll\_last},
this entail check challenges existing SL proof systems.
Additionally
for a proper reasoning, a proof system
also needs to
 infer the frame as the list \form{\seppred{\code{ll}}{y}} and a list segment from \form{t} to the node before
\form{t'}. Inferring such frame is nontrivial.
Our system generates the conjecture:
\savespace\[\savespace
\code{lemma ~c{:}~}
\form{\seppred{\code{ll\_last}}{t{,}t'}{\sep} \seppred{\code{ll}}{y}{\wedge}t{\neq}\nil} \rightarrow
\sepnode{\code{t'}}{c_1}{q} {\sep} \code{U_2}(t{,}t'\NI{,}q{,}y)
\]
Then, proves its validity as follows.
\begin{center}
\begin{small}
\def\defaultHypSeparation{\hskip 5pt}
\def\labelSpacing{0pt}

\AxiomC{(Base)\quad}
\AxiomC{\quad(Induction)}
\BinaryInfC{$\begin{array}{l}
\form{\seppred{\code{ll\_last}}{t{,}t'}{\sep} \seppred{\code{ll}}{y}{\wedge}t{\neq}\nil}
\form{{\ent_{\emptyset}}
\sepnode{\code{t'}}{c_1}{q} {\sep} \code{U_2}(t{,}t'\NI{,}q{,}y)
{\yields}(\horn_1{\wedge}\horn_2,\emp)
}
\end{array}$}
\DisplayProof
\end{small}\end{center}

\begin{center}
\begin{small}
\def\defaultHypSeparation{\hskip 1pt}
\def\labelSpacing{0pt}
\AxiomC{$\begin{array}{l}
\form{\emp}
\form{{\ent_{\{c\}}}
\emp {\yields} (\horn_1,\emp)
}
\end{array}$}
\LeftLabel{\scriptsize AF}
\UnaryInfC{$\begin{array}{l}
\form{\seppred{\code{ll}}{y}{\wedge}t'{=}t {\wedge}q{=}\nil{\wedge}t{\neq}\nil}
\form{{\ent_{\{c\}}}
\code{U_2}(t{,}t'\NI{,}q{,}y)
}
\end{array}$}
\LeftLabel{\scriptsize M}
\UnaryInfC{(Base):$\begin{array}{l}
\form{\sepnode{\code{t}}{c_1}{\nil} {\sep} \seppred{\code{ll}}{y} {\wedge}t'{=}t{\wedge}t{\neq}\nil}
\form{
{\ent_{\{c\}}}
\sepnode{\code{t'}}{c_1}{q} {\sep} \code{U_2}(t{,}t'\NI{,}q{,}y)
}
\end{array}$}
\DisplayProof
\end{small}\end{center}

\begin{center}
\begin{small}
\def\defaultHypSeparation{\hskip 1pt}
\def\labelSpacing{0pt}
\AxiomC{$\begin{array}{l}
\form{\emp}
\form{{\ent_{\{c\}}}
\emp {\yields} (\horn_1,\emp)
}
\end{array}$}
\LeftLabel{\scriptsize AF}
\UnaryInfC{$\begin{array}{l}
\form{
\sepnode{\code{t}}{c_1}{q_1} {\sep} \code{U_2}(q_1{,}t'\NI{,}q_2{,})
}
\form{{\ent_{\{c\}}}
 \code{U_2}(t{,}t'\NI{,}q_2{,}y)
}
\end{array}$}
\LeftLabel{\scriptsize M}
\UnaryInfC{$\begin{array}{l}
\form{\sepnode{\code{t}}{c_1}{q_1}{\sep}
\sepnode{\code{t'}}{c_1}{q_2} {\sep} \code{U_2}(q_1{,}t'\NI{,}q_2{,}y)
}
\form{{\ent_{\{c\}}}
\sepnode{\code{t'}}{c_1}{q} {\sep} \code{U_2}(t{,}t'\NI{,}q{,}y)
}
\end{array}$}
\LeftLabel{\scriptsize LAPP}
\UnaryInfC{(Induction):$\begin{array}{l}
\form{\sepnode{\code{t}}{c_1}{q_1} {\sep}\seppred{\code{ll\_last}}{q_1{,}t'}{\sep} \seppred{\code{ll}}{y}{\wedge}t{\neq}\nil}
\form{{\ent_{\{c\}}}
\sepnode{\code{t'}}{c_1}{q} {\sep} \code{U_2}(t{,}t'\NI{,}q{,})
}
\end{array}$}
\DisplayProof
\end{small}\end{center}
Relational assumptions are inferred as:
\savespace\[\savespace
\begin{array}{l}
\form{\horn_1{:}}~ \form{{\sep} \seppred{\code{ll}}{y}{\wedge}t'{=}t {\wedge}q{=}\nil{\wedge}t{\neq}\nil}~
 \form{~{\imply}~{\code{U_2}(t{,}t'{,}q{,}y)}}\\
\form{\horn_2{:}}~ \form{\sepnode{t}{\code{c_1}}{q_1}
{\sep}\code{U_2}(q_1{,}t'{,}q_2{,}y)}
 \form{~{\imply}~{\code{U_2}(t{,}t'{,}q_2{,}y)}}\\
\end{array}
\]
From \form{\horn_1} and \form{\horn_2}, our system
 synthesizes
the following definition for \code{U_2} as
\savespace\[\savespace
\begin{array}{l}
\form{\seppred{\code{U_2}}{\self{,}t'{,}q{,}y}} {\equiv}
 \form{\seppred{\code{ll}}{y}{\wedge}\code{\self}{=}t' {\wedge}q{=}\nil{\wedge}\self{\neq}\nil} 
 {\vee} \form{\exists q_1 {\cdot} \sepnode{\code{\self}}{c_1}{q_1} {\sep}
\seppred{\code{U_2}}{q_1{,}t'{,}q{,}y}}; \\
\end{array}
\]
Using theorem exploration presented in Sec \ref{sec:explore} and App. \ref{app.pred.trans},
our system generate the following two-way lemma to normalize the predicate \code{U_2}:
\savespace\[\savespace
\code{lemma ~conseq_0{:}}
\form{\seppred{\code{U_2}}{\self{,}t'{,}q{,}y} \leftrightarrow
\code{U_3}(\self{,}t'){\sep}\seppred{\code{ll}}{y}{\wedge}q{=}\nil{\wedge}\self{\neq}\nil}
\]
with \code{U_3} is a newly-inferred predicate as follows.
\savespace\[\savespace
\begin{array}{l}
\form{\seppred{\code{U_3}}{\self{,}t'}} {\equiv}
 \form{ \emp{\wedge}\code{\self}{=}t'} 
~ {\vee}~ \form{~\exists ~q_1 {\cdot}~ \sepnode{\code{\self}}{c_1}{q_1} {\sep}
\seppred{\code{U_3}}{q_1{,}t'}}; \\
\end{array}
\]
In summary,
our system successfully proves validity and infers frame
for the entailment.

We present a reasoning on both shape and size properties
of a more complicated revision of the \code{append}
method in App. \ref{mov.veri}.


 \subsection{Modular Verification with Incremental Specification Inference}\label{mov.veri}
 \begin{figure}
\begin{center} 
\begin{minipage}{0.99\textwidth} 
\begin{frameit} \savespace 
\begin{small}\savespace \[
\begin{array}{ll}
1 & \code{c_1{\sep}~append(int~n,  ~int~m)\{} \\
2 & \code{\quad c_1{\sep}x{=}creat\_ll(n);}  \\
3 & \code{\quad c_1{\sep}y{=}creat\_ll(m);}  \\
4 & \code{\quad c_1{\sep}t{=}x;}  \\
5 & \code{\quad while(t{\der}next)~ t{=}t{\der}next;} \\
6 & \code{\quad t{\der}next{=}y;} 
 \code{\quad return~x};\}
\end{array}
\qquad
\begin{array}{ll}
7 & \code{c_1{\sep}~create\_ll(int~s)\{} \\
8 & \code{~~if(s{=}0) ~return~\nil;}\\
9 & \code{~~else~\{}\\
10 & \code{\quad c_1{\sep}~ p{=}malloc(c_1);}\\
11 & \code{\quad p{\der}next{=}create\_ll(s{-}1);} \\
12 & \code{\quad return ~p}; 
 \code{\}} \\
\end{array}
\] 
\end{small}\savespace 
\end{frameit}
\end{minipage} \savespace \saveone
\caption{Code of method \code{append}.}
\label{fig.mov.append}
\end{center}\savespace \saveone
\end{figure}

In order to minimize the burden of
program verification, we pursue a compositional verification of programs whose
specifications are partially supplied.
Concretely, the user is only required to provide specifications
 of pre/post conditions
for critical methods and loop invariants whereas
specifications of the rest
 are inferred automatically.
This specification inference has been
implemented incrementally for the combined domains.
such that
 shape-only specifications are inferred first
 and then constraints over pure domains are additionally synthesized .
 In the context of heap-manipulating programs,
recursive methods and loop invariants normally relate
to recursive predicates; consequently, compositionally verifying these specifications
  requires both
 inductive reasoning and frame inference. Our proposed approach 
 brings the best support for
such verification.

\noindent To illustrate how our proof system is used to
{\em compositionally} and {\em incrementally} verify heap-manipulating programs,
consider the method \code{append}
in Fig. \ref{fig.mov.append} which appends a list pointed by \code{y}
to the end of a list pointed by \code{x}.
 The user provides the predicates \code{lln}, \code{lsegn} (Sec. \ref{sec:motivate})
and
\code{append}'s specification as:
\form{ \requires ~ n{>}0{\wedge}m{\geq}0 ~~ \ensures~ \seppred{\code{lln}}{\res{,}m{+}n}}.

Verifying safety properties that require both heap and data reasoning
has been studied in the literature, e.g.
 abstract interpretation \cite{Toubhans:VMCAI:2013},
TVLA \cite{Itzhaky:CAV:2014} and interpolation \cite{Albargouthi:ESOP:2015}.
Different to these proposals, ours 
is compositional and based on
the proposed inductive proof system.
 We enhance
inference techniques
 \cite{Brotherston-Gorogiannis:SAS:14,Loc:CAV:2014,Trinh:APLAS:2013} 
 to generate
specification and invariant for the combined domains of
 the method \code{create\_ll} and \code{while} loop  (of pre/post condition).
Our verification is bottom-up
 (i.e. verifying the method \code{create\_ll} and \code{while} loop before the method \code{append})
and incremental (i.e. analyzing {\em shape} and then {\em size} property
for the loop invariant inference).
Concretely, our system inferred the specification for  \code{create\_ll} and \code{while} loop as:
{\small  \form{{\bf\requires}~s{\geq}0 ~ {\bf\ensures} ~ \seppred{\code{lln}}{\res{,}s}}} and
{\small \code{(s_3)} \form{{\bf\requires} ~ \exists q{\cdot}\sepnode{\code{t}}{c_1}{q} {\sep} 
 \seppred{\code{lln}}{q{,}i}{\wedge}i{\geq}0 ~ {\bf\ensures} ~ \seppred{\code{lsegn}}{t{,}t'{,}j} {\sep} \sepnode{\code{t'}}{c_1}{\nil} {\wedge}{j{=}i}}}, resp.

In the following, we present the specification inference 
 for the loop.
To infer shape specification of the loop invariant
the system initially
 generates specification with two unknown shape predicates
as follows: 
 \form{\requires ~ \code{U_{pre}}(t)} \form{ \ensures ~ \code{U_{post}}(t,t')},
whereas \form{t'} is the value of \form{t} after the loop.
Using the {modular} shape analysis in \cite{Loc:CAV:2014},
the predicate \code{U_{pre}} and \code{U_{post}} are synthesized as
\savespace\[\savespace 
\begin{array}{ll}
 \seppred{\code{U_{pre}}}{\self} & {\equiv}~ \exists q{\cdot}\sepnode{\code{\self}}{c_1}{q} {\sep} 
 \seppred{\code{U_{1}}}{q} \\
\form{\seppred{\code{U_1}}{\self}} & \equiv~
\form{\emp} {\wedge} \form{\code{\self}{=}\nil} ~~ {\vee} \form{~\exists ~q {\cdot}~ \sepnode{\code{\self}}{c_1}{q} {\sep} \seppred{\code{U_1}}{q} }\\
\form{\seppred{\code{U_{post}}}{\self{,}l}} & \equiv~
\form{\sepnode{\code{\self}}{c_1}{\nil}} {\wedge} \form{\code{\self}{=}l} ~~ {\vee} \form{~\exists ~q {\cdot}~ \sepnode{\code{\self}}{c_1}{q} {\sep} \seppred{\code{U_{post}}}{q{,}l} {\wedge}\self{\neq}l} \\
\end{array}
\]
whereas \code{U_1} is an auxiliary predicate.
Whenever receiving these predicate definitions,
 our theorem exploration
component will generate new lemmas
to study interesting properties of these predicates.
In this example, our system generates the following lemmas
to explicate the separation between two parameters
of the predicate \code{U_{post}}:
\savespace\[\savespace
\begin{array}{l}
 \seppred{\code{lemma~consep~} \code{U_{post}}}{\self{,}l} ~ {\leftrightarrow}~  
 \seppred{\code{U_{2}}}{\self{,}l} {\sep} \sepnode{\code{l}}{c_1}{\nil}\\
\form{\seppred{\code{U_2}}{\self{,}s}} ~ {\equiv}~
\form{\emp} {\wedge} \form{\code{\self}{=}s} ~~ {\vee} \form{~\exists ~q {\cdot}~ \sepnode{\code{\self}}{c_1}{q} {\sep} \seppred{\code{U_2}}{q{,}s} {\wedge}\self{\neq}s}
\end{array}
\]
Then, the shape invariant of the loop is constructed as follows: 
\savespace\[\savespace \form{\requires ~ \exists q{\cdot}\sepnode{\code{x}}{c_1}{q} {\sep} 
 \seppred{\code{U_{1}}}{q} \quad \ensures ~ \seppred{\code{U_{2}}}{x{,}x'} {\sep} \sepnode{\code{x'}}{c_1}{\nil}}
\]
To extend the above shape invariant with the size property,
we first use predicate extension \cite{Trinh:APLAS:2013}
to automatically
 append the size property into \code{U_1} (\code{Un_1})
 and \code{U_2} (\code{Un_2})\savespace
\[\savespace
\begin{array}{ll}
 \form{\seppred{\code{Un_1}}{\self{,}n}} & \equiv~
\form{\emp} {\wedge} \form{\code{\self}{=}\nil{\wedge}n{=}0} ~~ {\vee} \form{~\exists ~q {\cdot}~ \sepnode{\code{\self}}{c_1}{q} {\sep} \seppred{\code{Un_1}}{q{,}n{-}1} }\\
\form{\seppred{\code{Un_2}}{\self{,}s{,}n}} & \equiv~
\form{\emp} {\wedge} \form{\code{\self}{=}s{\wedge}n{=}0} ~~ {\vee} \form{~\exists ~q {\cdot}~ \sepnode{\code{\self}}{c_1}{q} {\sep} \seppred{\code{Un_2}}{q{,}s{,}n{-}1}{\wedge}\self{\neq}s }
\end{array}
\]
Our theorem exploration, again, generates
two lemmas 
 to match \code{Un_1}
with \code{lln} and \code{Un_2} with \code{lsegn}
as: \form{\seppred{\code{Un_1}}{\self{,}n} {\leftrightarrow} \seppred{\code{lln}}{\self{,}n}} and \form{\seppred{\code{Un_2}}{\self{,}s{,}n} {\leftrightarrow} \seppred{\code{lsegn}}{\self{,}s{,}n}}.
After that,
we generate specification with unknown pure predicates \code{P_2}, \code{P_3}:
\savespace\[\savespace
\form{\requires ~ \exists q{\cdot}\sepnode{\code{t}}{c_1}{q} {\sep} 
 \seppred{\code{lln}}{q{,}i}{\wedge}\code{P_2}(i) ~~ {\ensures} ~ \seppred{\code{lsegn}}{t{,}t'{,}j} {\sep} \sepnode{\code{t'}}{c_1}{\nil} {\wedge}\seppred{\code{P_3}}{i{,}j}}
\]whereas \code{P_2} and \code{P_3} are placeholders to capture constraints
over the {\em size} variables.
Using {\sobd} for pure properties inference \cite{Trinh:APLAS:2013},
the following definitions are synthesized:
\form{\seppred{\code{P_2}}{i}{\equiv}i{\geq}0}
and \form{\seppred{\code{P_3}}{i{,}j}{\equiv}j{=}i}.
Finally,
loop invariant is inferred as the specification \code{s_3}.


\noindent This loop invariant is now used in the verification of
 the main method \code{append}.
To verify the correctness and memory safety (no null-dereference and no leakage) of \code{append}, our system generates
 and successfully proves
the following three verification conditions: \footnote{For simplicity, we discard the verification conditions at line 2 and line 3.}
\begin{small}\savespace
\[\savespace
\begin{array}{ll}
\code{VC_1}. & \seppred{\code{lln}}{x{,}n}{\sep}\seppred{\code{lln}}{y{,}m} {\wedge}n{>}0{\wedge}m{\geq}0
~~{\ent}_{\lemstore}~~  {\exists} q{\cdot}\sepnode{\code{t}}{c_1}{q} {\sep} 
 \seppred{\code{lln}}{q{,}i}{\wedge}i{\geq}0 \\
& ~ {\yields}
 (\true{,}\seppred{\code{lln}}{y{,}m}{\wedge}x{=}t{\wedge}t{\neq}\nil{\wedge}n{=}i{-}1{\wedge}n{>}0
{\wedge}i{\geq}0) ~~  \text{// line 5, pre- proving (of loop)}\\
\code{VC_2}. & \seppred{\code{lsegn}}{t{,}t'{,}j}{\sep}\sepnode{\code{t'}}{c_1}{\nil} 
{\sep}\seppred{\code{lln}}{y{,}m} {\wedge}{i{=}j} {\wedge}x{\neq}\nil{\wedge}n{=}i{-}1{\wedge}n{>}0 ~~{\ent}_{\lemstore}~~  \sepnode{\code{t'}}{c_1}{\_}\\
& ~{\yields}
(\true{,}\seppred{\code{lsegn}}{t{,}t'{,}j}{\sep} \seppred{\code{lln}}{y{,}m} {\wedge}{i{=}j} {\wedge}x{=}t{\wedge}t{\neq}\nil{\wedge}n{=}i{-}1{\wedge}n{>}0){\wedge}m{\geq}0 \\
 &
  \qquad \qquad\text{// before line 6, no null-dereference} \\
\code{VC_3}. & \seppred{\code{lsegn}}{\res{,}t'{,}j}{\sep}\sepnode{\code{t'}}{c_1}{y} 
{\sep}\seppred{\code{lln}}{y{,}m} {\wedge}{i{=}j}{\wedge}x{=}t {\wedge}t{\neq}\nil{\wedge}n{=}i{-}1{\wedge}n{>}0{\wedge}m{\geq}0  \\
& ~~{\ent}_{\lemstore}~~ \seppred{\code{lln}}{\res{,}m{+}n} ~  {\yields}
(\true{,}\textcolor{blue}{\emp}{\wedge}x{=}t{\wedge}t{\neq}\nil{\wedge}n{=}i{-}1{\wedge}n{>}0
{\wedge}i{\geq}0) \\
 & \qquad \qquad \text{// after line 6, post-condition, \textcolor{blue}{no leakage}}
\end{array}
\]
\end{small}
whereas \form{\lemstore} is the set of lemmas either supplied by the user
or generated by our system (i.e. to explore predicate relations like the \code{conseq} lemmas above).

\noindent We highlight two advantages achieved from our proposed approach.
First, if the two-way lemma \code{consep} was not synthesized,
the condition \code{VC_2} would be generated as: \form{\code{U_{post}}({t{,}t'}){\sep}\seppred{\code{lln}}{y{,}m} {\wedge}x{=}t{\wedge}{i{=}j} {\wedge}t{\neq}\nil{\wedge}n{=}i{-}1{\wedge}n{>}0 ~{\ent}_{\lemstore}~  \sepnode{\code{t'}}{c_1}{\_}}.
We are not aware of any proof systems that are capable of
discharging such obligation and inferring the residual frame, simultaneously.
Second, simultaneously automated proving \code{VC_3} and inferring its residual heap \form{\emp}
(to confirm no memory is leaked)
require nontrivial inductive reasoning which is not supported
by most existing SL entailment procedures.


\end{document}